\documentclass{article}

\usepackage{jheppub}
\usepackage[utf8]{inputenc}
\usepackage{multirow}
\usepackage[toc,page]{appendix}
\usepackage[inline]{enumitem}
\usepackage{booktabs}
\usepackage{amsmath}
\usepackage{amsfonts}
\usepackage{amssymb}
\usepackage{graphicx}
\usepackage{ulem} 
\usepackage{slashed}
\usepackage{hyperref}
\usepackage{float}
\usepackage{csquotes}
\usepackage[inline]{enumitem}
\usepackage{graphicx}
\usepackage{caption}
\usepackage{subcaption}
\usepackage{diagbox}
\usepackage[font=small,labelfont=bf]{caption} % Optional: For smaller captions
\usepackage{tabularx}
\usepackage[ruled,vlined]{algorithm2e}
\usepackage[most]{tcolorbox}

\usepackage{placeins}
\usepackage{booktabs}

\makeatletter
\gdef\@fpheader{\vline}
\makeatother

\newcolumntype{C}[1]{%
  >{%
    \hsize=\dimexpr\calc{1/#1}\hsize\relax%
    \linewidth=\dimexpr\hsize\relax%
    \centering\arraybackslash%
  }X%
}

\author[a,b]{Ambre Visive}
\emailAdd{avisive@nikhef.nl}

\author[c]{Roberto Ruiz de Austri}
\emailAdd{rruiz@ific.uv.es}

\author[d,e]{Polina Moskvitina}
\emailAdd{p.moskvitina@nikhef.nl}

\author[a,b]{Clara Nellist}
\emailAdd{c.nellist@nikhef.nl}

\author[d,e]{Sascha Caron}
\emailAdd{scaron@nikhef.nl}

\title{Masked-Token Prediction for Anomaly Detection at the Large Hadron Collider}

\abstract{
Anomaly detection in High Energy Physics requires identifying rare signals against overwhelming backgrounds, without prior knowledge of the signal. We present the first application of masked-token prediction, a technique from Large Language Models, to this problem. A lightweight encoder architecture trained solely on background events captures the structure of Standard Model (SM) physics; at inference, sequences deviating from this learned structure are flagged as anomalous.
We evaluate the approach on searches for four-top-quark production and supersymmetric gluino pair production, both featuring top-rich final states with substantial missing transverse energy, covering SM and beyond the Standard Model (BSM) scenarios. Strong performance on the four-top signature, which closely resembles background, demonstrates the method's sensitivity to subtle deviations.
We further show that the tokenization strategy significantly impacts performance: deep-learned tokenization via vector-quantized variational autoencoders (VQ-VAE) outperforms look-up table tokenization. Comparison with established anomaly detection baselines confirms robustness.
These results highlight the potential of token-based collider data representations combined with transformer architectures for new-physics discovery. Once trained on SM background, the model transfers across different BSM searches, enabling scalable, model-independent anomaly detection at reduced computational cost.

}

\begin{document}

\maketitle

\clearpage

%%%%%%%%%%%%%%%%%%%%%%%
\section{Introduction}
%%%%%%%%%%%%%%%%%%%%%%

Recent advances in machine learning (ML) have led to major improvements across a wide range of tasks and fields. Since the introduction of the first contemporary Large Language Models (LLMs)~\cite{OpenAi2017llm}, these architectures have been widely adopted for applications such as text generation and translation \cite{minaee2025largelanguagemodelssurvey}. While all transformer-based models benefit from the ability to capture long range dependencies through the self-attention mechanism, introduced in Ref.~\cite{DBLP:journals/corr/VaswaniSPUJGKP17}, the remarkable success of LLMs in natural language processing is largely attributed to the availability of extremely large training datasets  \cite{stochasticparrotsBender2021}, made possible by the inherent richness and ubiquity of human language, which offers a scale and diversity of examples that is rarely matched in other domains.

With the unprecedented volume of data expected from the current and upcoming runs of the Large Hadron Collider (LHC) \cite{atlas2025highlightshllhcphysicsprojections}, similar techniques may be leveraged to address some of the key challenges in high-energy physics. In this work, we propose the use of LLM-inspired models as unsupervised anomaly detectors for particle physics data. Using the masked-token prediction strategy introduced by BERT \cite{devlin2019bertpretrainingdeepbidirectional}, our model learns the underlying physics of the background processes during training from the context provided by the remaining particles in each event. Deviations from the learned physics patterns, quantified through reconstruction scores at inference, serve as indicators of potential signal events, enabling detection without prior knowledge of the signal. 

%\textbf{SC: I would not cite conference work like this.}
%\textbf{CN: Also, we might not want to highlight that it's been `published' elsewhere.}
%Building on Refs.~\cite{Visive:2025flm,Visive:2026OJ}, which first introduced .
Our method provides a detailed proof of concept and examines how different strategies for representing collision events through tokenization impact the model performance. We demonstrate that this approach has the potential to enhance sensitivity to rare Standard Model (SM) processes, such as the four-top-quark simultaneous production process, and to uncover signatures of physics Beyond the Standard Model (BSM) with modest computing resources.
Masked-token prediction provides a natural mechanism for anomaly detection within masking-based foundation models or even Large Physics Models \cite{Golling:2024abg,Barman:2025wfb}, since the same pretraining objective used to learn background structure can directly serve as an anomaly score at inference.

The remainder of this paper is organized as follows. In Section~\ref{sec:dataset}, we describe the physics benchmarks considered in this work %, the Dark Machines datasets, 
and the event representation used throughout the analysis. In Section~\ref{sec:models}, we introduce the ML framework, including the masked-token prediction strategy for anomaly detection and the vector-quantized variational autoencoder (VQ-VAE) model used for learned tokenization. In Section~\ref{sec:Tokenization}, we present the different tokenization approaches explored in this study, namely the look-up table (LUT) and VQ-VAE-based schemes. In Section~\ref{sec:Results}, we report the corresponding results, including the comparison between tokenization strategies and with established unsupervised anomaly detection methods. Finally, in Section~\ref{sec:Conclusions}, we summarize our conclusions and discuss the main implications of this work. 

%\RR{I think it's better to avoid the use of LLM-like since it is not precise. Use LLMs-inspired in introduction, abstract and conclusions. Then in technical sections we can use masked-token transformer model and when we  talk about the training objective we can refer to BERT-inspired.} \AV{Good for me. I'll remove it in the rest of the text.}

%%%%%%%%%%%%%%%%%%%%%%%%%%%%%%%%%%%%%%%%%%
\section{Dataset} \label{sec:dataset}
%%%%%%%%%%%%%%%%%%%%%%%%%%%%%%%%%%%%%%%%%%%
%In this section, the datasets used in this study are described, outlining their physical motivation and generation specifics.
%%%%%%%%%%%%%%%%%%%%%%%%%%%%%%%
\subsection{Physics Motivation} 
\label{sec:physics}
%%%%%%%%%%%%%%%%%%%%%%%%%%%%%%%%%%%%%%%

%%%%%%%%%%%%%%%%%%%%%%%%%%%%%%%
\subsubsection{Four-top-quark events} 
\label{sec:4topdata}

The simultaneous production of four top quarks is among the rarest SM process that can be measured with the current LHC dataset. Owing to its very small cross section and highly complex final state, it provides a particularly demanding benchmark for anomaly detection methods. Events in this channel exhibit high object multiplicity and non-trivial correlations among reconstructed objects, making this process especially suitable for testing whether unsupervised methods can identify subtle SM signals.

%Each top quark decays almost exclusively through $t \to Wb$, since $|V_{tb}|^2 \approx 1$ \PM{I think it would be better to add a reference here, such as the PDG}. Depending on the decay modes of the $W$ bosons, four-top events may contain between 0 and 4 leptons and between 4 and 12 jets, including at least four $b$-jets.

%Since each top quark decays almost exclusively into $Wb$, four-top events may contain between 0 and 4 leptons and between 4 and 12 jets, depending on the decay modes of the $W$ bosons, including at least four $b$-jets.

Since each top quark decays almost exclusively via \(t\rightarrow W+b\), due to \(|V_{tb}|^2\sim{}1\), four-top events may contain between 0 and 4 leptons and between 4 and 12 jets, depending on the decay modes of the $W$ bosons, including four $b$-jets.

Because of the strong similarity between the four-top topology and other high-multiplicity SM final states, this process constitutes a challenging and well-motivated benchmark for unsupervised searches. %For the present study, we adopt the four-top benchmark considered in Ref.~\cite{SciPostPhys.19.1.028}. 
\(t\bar{t}t\bar{t}\) is the \textbf{SM signal} process to extract, while \(t\bar{t}W^+W^-\), \(t\bar{t}W^\pm\), \(t\bar{t}Z\) and \(t\bar{t}H\) constitute the \textbf{4-top background}.

%%%%%%%%%%%%%%%%%%%%%%%%%%%%%%%
\subsubsection{SUSY gluino-gluino} 
\label{sec:BSMdata}
%%%%%%%%%%%%%%%%%%%%%%%%%%%%%%%%%%%%%%%
%To evaluate the performance of our method on BSM processes, we consider a supersymmetry (SUSY) scenario. With a highly complex, high-multiplicity event topologies, the simultaneous production of SUSY-gluino pairs is the BSM benchmark of our study. As each gluino decays into a pair of top quark and a neutralino ($\tilde{g}\rightarrow \bar{t}+\bar{t}$, $\bar{t}\rightarrow t+\tilde{\chi}^0$ \cite{berger2011boostedtopsgluinodecays}), the $\tilde{g}\tilde{g}$  process yield a final state similar to the one of the four-top-quarks processes (described in details in Section~\ref{sec:4topdata}) together with a substantial missing transverse energy from the neutralinos. 

%As a consequence, \(t\bar{t}t\bar{t}\) events become part of the background for this BSM scenario, in addition to the other top processes that share similarities with the four-top signature (\(t\bar{t}WW\), \(t\bar{t}W\), \(t\bar{t}Z\) and \(t\bar{t}H\)). 
%In this part of the study, the $\tilde{g}\tilde{g}$ events are treated as the \textit{BSM signal}, while the aforementioned SM processes (\(t\bar{t}WW\), \(t\bar{t}W\), \(t\bar{t}Z\), \(t\bar{t}H\) and \(t\bar{t}t\bar{t}\)) form the \textit{SM background} from which the BSM signal has to be extracted.

As a benchmark BSM, we consider a supersymmetric (SUSY) gluino-pair production scenario. %based on the Dark Machines dataset~\cite{Aarrestad_2022}. 
In the sample used in this work, the gluino mass is fixed to $m_{\tilde g}=1~\mathrm{TeV}$ and the neutralino mass to $m_{\tilde\chi^0_1}=1~\mathrm{GeV}$. The gluino is assumed to decay via an off-shell stop according to
\[
\tilde g \rightarrow t \bar t \tilde\chi^0_1 \, ,
\]
so that gluino-pair production yields final states with multiple top quarks and substantial missing transverse energy. We treat the gluino-pair events as \textbf{BSM signal} and \textbf{SM background} contains the following SM processes : $jj$, $l^\pm\nu_l$, $\gamma j$, $l^+l^-$, $t\bar{t}$, $tj$, $\bar{t}j$, $W^+W^-$, $tW^\pm$, $\bar{t}W^\pm$, $\gamma\gamma$, $W^\pm\gamma$, $ZW^\pm$, $Z\gamma$, $ZZ$, $H$, $t\bar{t}\gamma$, $t\bar{t}Z$, $t\bar{t}H$, $\bar{t}\gamma$, $t\bar{t}W^\pm$, $tZ$, $\bar{t}Z$, $t\bar{t}t\bar{t}$, $t\bar{t}W^+W^-$.

%and the corresponding SM sample (see Table~2 of Ref.~\cite{Aarrestad_2022}) as \textbf{SM background}.

%%%%%%%%%%%%%%%%%%%%%%%%%%%%%%%
%\subsection{The Dark Machines datasets} 
\subsection{Event samples and representation}
\label{sec:DMdata}

\begin{table}[b] %htbp
\centering
\begin{minipage}[t]{0.55\textwidth}
\centering
\caption{Mapping between the Standard Model processes and their associated labels for the $t\bar{t}t\bar{t}$ study.}\label{table:processlabels}
\begin{tabular}{ l l c } 
\toprule
\textbf{Benchmark} & \textbf{Process}& \textbf{{\fontfamily{cmtt}\selectfont process ID }}\\ 
\hline
\multirow{5}{*}{$t\bar{t}t\bar{t}$ study}
&\(tttt\) & 1\\ 
&\(ttH\) & 2\\ 
&\(ttW\) & 3\\ 
&\(ttWW\) & 4\\ 
&\(ttZ\) & 5\\ 
\addlinespace
\midrule
\addlinespace
\multirow{2}{*}{BSM study}
& SUSY \(g\tilde{g}\) & 1\\ 
& SM background & 0\\ 
\bottomrule
\end{tabular}
% \vspace{1em}
% \caption{Mapping between the processes and their associated labels for the BSM study.}\label{table:SMlabels}
% \begin{tabular}{ c c } 
% \toprule
% \textbf{SM process }& \textbf{{\fontfamily{cmtt}\selectfont process ID }}\\ 
% \hline
% \bottomrule
% \end{tabular}
\end{minipage}
\hfill
\begin{minipage}[t]{0.40\textwidth}
\centering
\caption{Particles with their associated symbols and tags}\label{table:particle-object}
\begin{tabular}{ c c c } 
\toprule
\textbf{Object} & \textbf{{\fontfamily{cmtt}\selectfont Symbol ID}} & \textbf{{\fontfamily{cmtt}\selectfont tag}}\\ 
\hline%\hline
jet & {\fontfamily{cmtt}\selectfont j} & 1\\ 
%\hline
b-tagged jet & {\fontfamily{cmtt}\selectfont b} & 2\\  
%\hline
positron & {\fontfamily{cmtt}\selectfont e+} & 3\\ 
%\hline
electron & {\fontfamily{cmtt}\selectfont e-} & 4\\ 
%\hline
anti-muon & {\fontfamily{cmtt}\selectfont mu+} & 5\\ 
%\hline
muon & {\fontfamily{cmtt}\selectfont mu-} & 6\\ 
%\hline
photon & {\fontfamily{cmtt}\selectfont g} & 7\\
\bottomrule
\end{tabular}
\end{minipage}
\end{table}

The analyses presented in this work are based on benchmark collider event samples corresponding to the two physics scenarios described in Section~\ref{sec:physics}. For the four-top study, we adopt the benchmark setup of Ref.~\cite{SciPostPhys.19.1.028}, while for the BSM study we use the Dark Machines channel~2b benchmark~\cite{Aarrestad_2022,Caron:2022wrw,Caron:2021wmq} with only the gluino-pair events as signal and the associated SM sample as background (see Table~2 of Ref.~\cite{Aarrestad_2022}). %\PM{It's somewhat repetitive of Sections~2.1.1 and~2.1.2.}. 
Detailed descriptions of the underlying event generation, detector simulation, and original data format are provided in the original references and are not repeated here. %In both cases, the events are subsequently transformed into the sequence representation used by our models.

From the original data format, the {\fontfamily{cmtt}\selectfont process ID} is extracted and mapped to an integer corresponding the process class (Tables~\ref{table:processlabels}) to create labeled datasets. Each event is then represented as a sequence consisting of: particle-object type and charge tags (see Table~\ref{table:particle-object}) ordered by object type and decreasing $p_T$, the $\|E^{miss}_T\|$ and $\phi_{E^{miss}_T}$ of the event, and the four-momenta of all particle-objects in the same ordering, Equation \ref{eq:DMdatastep2}. Sequences are padded with zeros to a fixed size corresponding to 18 particle-objects, in order to obtain a uniform input dimensionality. The treatment of padded entries by the masked-token prediction models is described in Sections~\ref{sec:VQVAE} and~\ref{sec:Tokenization}. Finally, the resulting datasets are split into training (80\%), validation (10\%), and test (10\%) subsets.
\vspace{-0.1cm}
\begin{equation}
obj_1;  obj_2; ...; obj_{18}; \|E^{miss}_T\|; \phi_{E^{miss}_T}; E_1; p_{T,1}, \eta_1; \phi_1; E_2; p_{T,2}, \eta_2; \phi_2;...;  E_{18}; p_{T,18}, \eta_{18}; \phi_{18} 
\label{eq:DMdatastep2}
\end{equation}

%Each event is encoded as a fixed-length sequence containing global event information together with reconstructed particle-object features. Since the number of reconstructed objects varies from event to event, sequences are padded to a common length of 18 particles in order to obtain a uniform input dimensionality. The treatment of padded entries in the tokenization models is described in Sections~\ref{sec:VQVAE} and~\ref{sec:Tokenization}. Unless stated otherwise, the datasets are split into training, validation, and test subsets following the setup adopted in this work. 

The masked-token prediction model that we use in this work takes fixed-length sequence of tokens as input (see Section~\ref{sec:LLM}). Constructing these sequences from the previously described datasets requires a dedicated tokenization procedure, which is described in detail in Section~\ref{sec:Tokenization} including the encoding schemes, alternative representations, and their respective performances (see Section~\ref{sec:performance}).

%Each collision event is represented as a fixed length sequence of tokens suitable for processing by the model used for anomaly detection, an LLM-inspired architecture described in Section~\ref{sec:LLM}. The sequences encode both global event information and per-particle features. Individual tokens correspond either to the missing transverse momentum, its azimuthal angle, or the reconstructed particle objects, characterized by its tag, transverse momenta, pseudo-rapidity and azimuthal angle ($obj; p_{T}, \eta; \phi$). To construct these sequences from the Dark Machines dataset (Section~\ref{sec:DMdata}), a dedicated tokenization procedure is required. The detailed tokenization procedures explored in this work, including the specific encoding schemes, alternative representations and their performances, are discussed in Section~\ref{sec:Tokenization}. 

%\noindent A detailed description of the Dark Machines dataset preprocessing pipeline is available in Ref.~\cite{SciPostPhys.19.1.028} and Ref.~\cite{Moskvitina:2025epjc}, as well as in Ref.~\cite{Aarrestad_2022}.

%%%%%%%%%%%%%%%%%%%%%%%%%%%%%%%
%\subsection{Data Format} 
%\label{sec:dataformat}
%%%%%%%%%%%%%%%%%%%%%%%%%%%%%%%%%%%%%%%

%%%%%%%%%%%%%%%%%%%%%%%%%%%%%%%%%%%%%%%%%%%%%%%%%%%%%
\section{Machine Learning Models} %consider Architectures instead of Models
\label{sec:models}
%%%%%%%%%%%%%%%%%%%%%%%%%%%%%%%%%%%%%%%%%%%%%%%%%%%%%%

%%%%%%%%%%%%%%%%%%%%%%%%%%%%%%%%%
\subsection{Masked-Token Prediction models} %and LLM-inspired models} 
\label{sec:LLM}
%%%%%%%%%%%%%%%%%%%%%%%%%%%%%%%%%%

Transformer-based architectures, which form the backbone of modern LLMs, owe much of their success to the self-attention mechanism~\cite{DBLP:journals/corr/VaswaniSPUJGKP17}, which allows them to capture contextual relationships between elements of an input sequence regardless of their relative positions.
%Transformer based architectures, commonly referred to as LLMs when the model’s inputs are sequences of discrete units known as \textit{tokens} (originally designed to represent words or subword fragments in natural language), owes most of their success from the self attention mechanism \cite{DBLP:journals/corr/VaswaniSPUJGKP17}, which enables the model to capture contextual relationships between tokens regardless of their relative positions. 
Their ability to model long range dependencies makes them naturally suited for representing particle physics events, where correlations between widely separated objects often carry important physical information. In this work, only the encoder component of a lightweight transformer architecture is used to process collider events represented as token sequences. The model begins with an embedding layer that maps each token to a dense vector representation. 
% Positional information may be incorporated through a variety of positional encoding schemes (sinusoidal, learned, rotary or convolutional encodings, etc.) but our studies do not show a clear preference among these choices for the tasks considered. 
Positional information may be incorporated through a variety of encoding schemes, including sinusoidal and learned positional encodings~\cite{DBLP:journals/corr/VaswaniSPUJGKP17}, rotary positional embeddings~\cite{su2023roformerenhancedtransformerrotary}, and convolution-based positional encodings~\cite{chu2023conditionalpositionalencodingsvision}, but our studies do not show a clear preference among these choices for the tasks considered.
The core of the network consists of two transformer encoder layers, each equipped with four self attention heads, allowing the model to capture contextual relationships between tokens, regardless of their relative positions. The final hidden representation is passed through a linear projection layer to map it to the desired number of output classes, followed by a softmax activation layer to convert these raw scores into probabilities.

This architecture is trained using a masked token prediction objective inspired by BERT %\cite{devlin2019bertpretrainingdeepbidirectional}
, enabling unsupervised learning of the dominant background structure. The corresponding anomaly detection strategy is described in Section~\ref{sec:LLManomalydetector}.

A crucial prerequisite for applying masked-token transformer models to scientific data is the construction of an appropriate token representation, referred to as the \textit{tokenization}, which enables the model to interpret complex, human meaningful information by converting it into a format suitable for the model. In this work, each particle physics event, either recorded by detectors or generated through simulation, is  a sequence of tokens. Each token encodes relevant physical quantities such as the particle-type and charge, the transverse momentum, the pseudorapidity, the azimuthal angle, or the missing transverse energy %\PM{It feels like this point has already been mentioned several times by this stage of the paper.}. 
%\AV{Not anymore normally, removed it from above.}
This will be discussed in more detailed in Section \ref{sec:Tokenization}.

%\vspace{1cm}
\begin{figure}[htbp]
    \centering
    \includegraphics[width=0.95\linewidth]{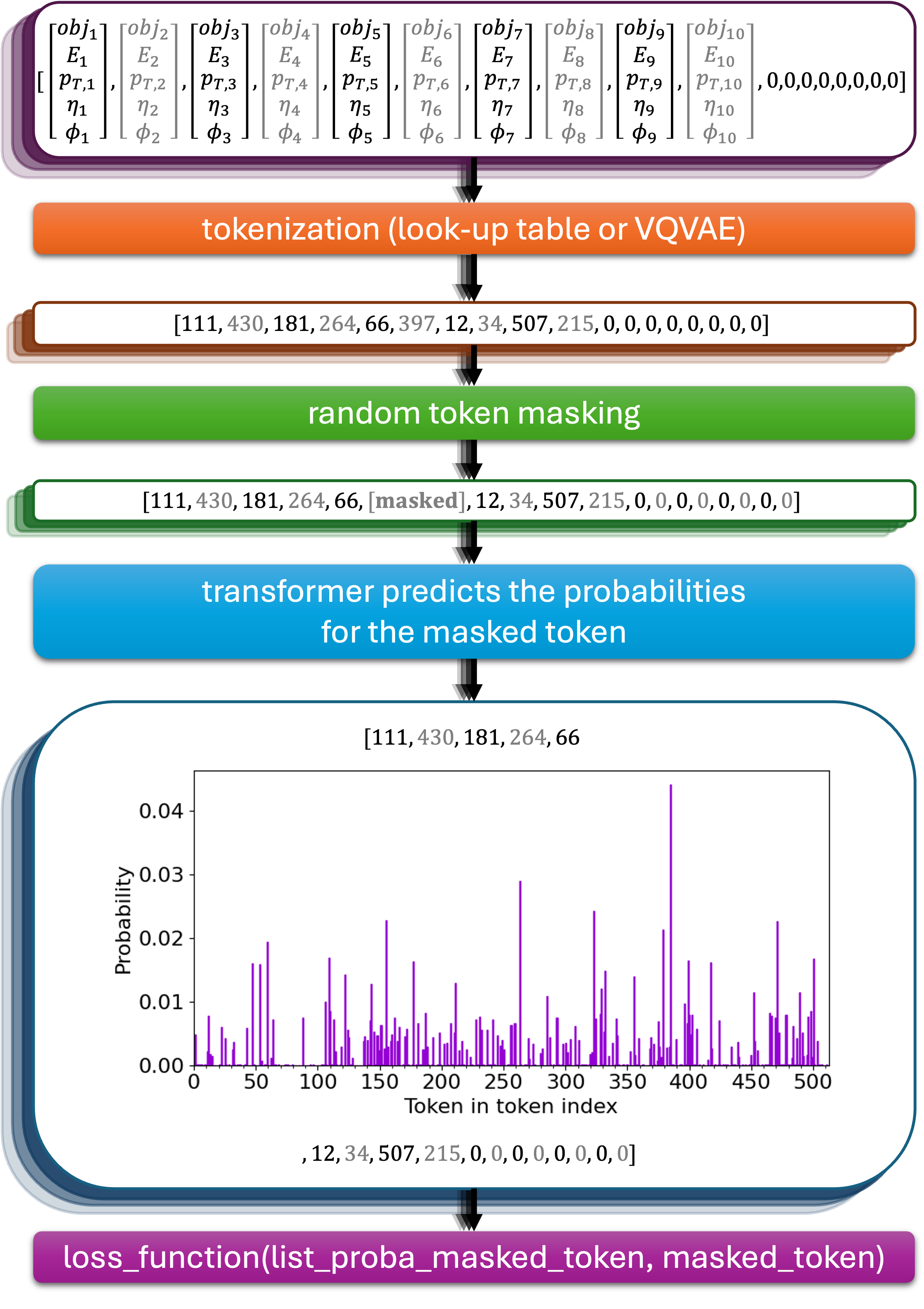}
    \caption{Visualisation of the procedure for a random event where the token of index 5 ($397$) is masked.}
    \label{fig:model}
\end{figure}

%%%%%%%%%%%%%%%%%%%%%%%%%%%%%%%%%
\subsection{Strategy for anomaly detection} 
\label{sec:LLManomalydetector}
%%%%%%%%%%%%%%%%%%%%%%%%%%%%%%%%%%

The central idea of this work is to train the model on background events only, so that deviations from the learned structure can be identified as anomalies during inference. To this end, the encoder component of our lightweight transformer architecture is coupled to a masked token prediction strategy. %, following the concept introduced by BERT \cite{devlin2019bertpretrainingdeepbidirectional}. 

Following tokenization, the model is trained exclusively on background events. During training and validation, for each event in a batch, a single token within each event sequence is randomly masked, and the model is tasked with predicting the original token from the surrounding context. The output of the final projection layer is a probability distribution over the token vocabulary (probability vector indicating the likelihood of each token being the correct reconstruction), and training proceeds by minimizing the sparse categorical cross entropy between the predicted distribution and the true masked token (see Figure~\ref{fig:model}). Model parameters are updated using the Adam optimizer~\cite{kingma2017adam}, 
%and training is stopped early if the validation loss does not improve by at least 0.1\% for five consecutive epochs to avoid overfitting. 
with early stopping applied to prevent overfitting.
Through this procedure, the model learns to reconstruct masked tokens accurately for events consistent with the background distribution. Conversely, events originating from processes not encountered during training are expected to yield poorer reconstruction performance, enabling their identification as anomalies.

\begin{figure}[h!]
    \centering
    \includegraphics[width=0.6\linewidth]{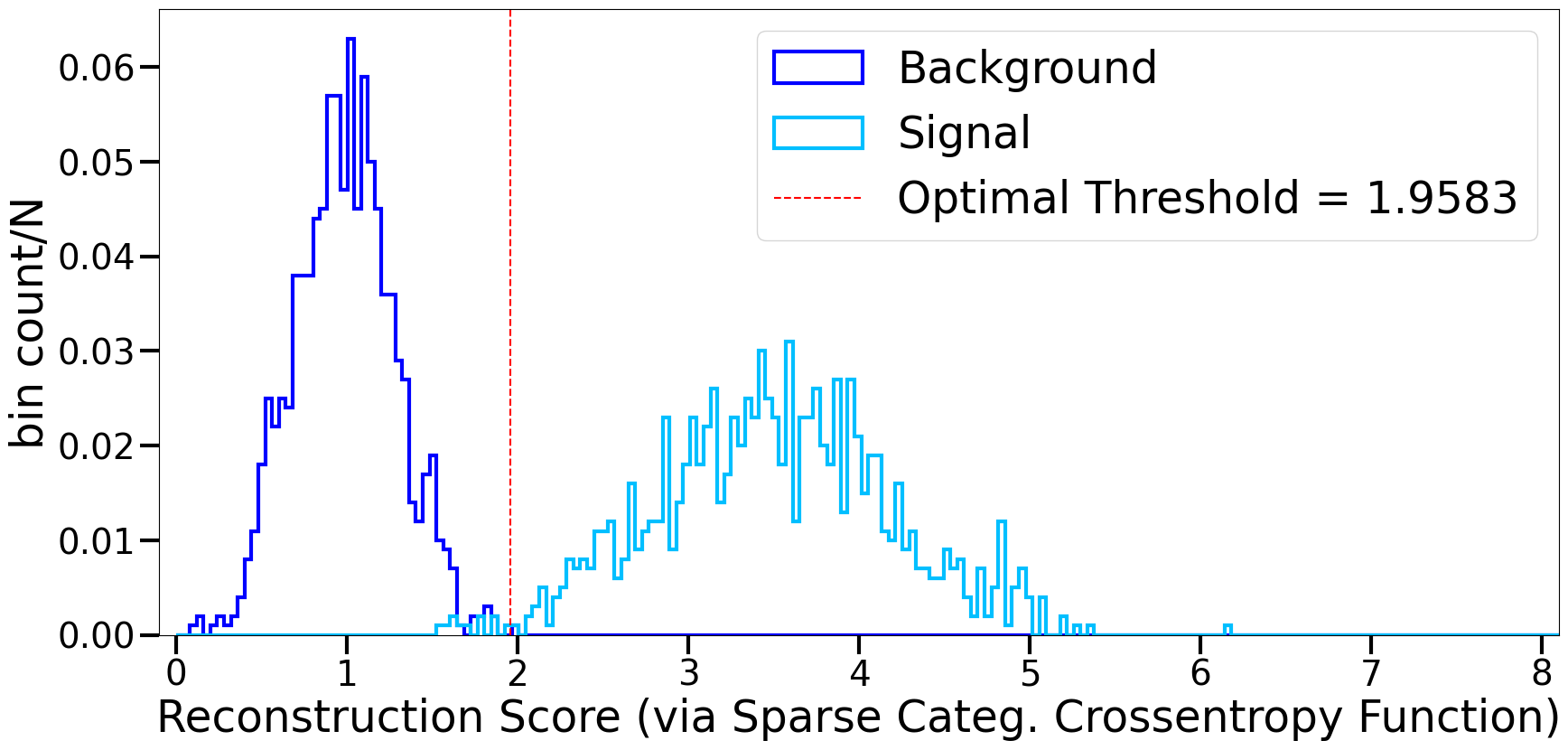}
    \caption{Illustrative distribution of anomaly scores in an ideal scenario.}
    \label{fig:perfectHisto}
\end{figure}

During inference, both background and signal events are processed by the trained model. For each event, every token is masked and reconstructed one at a time, resulting in a set of reconstruction losses across the sequence. These per-token losses are then averaged %(\SC{How ?} 
to produce a normalised single event-level reconstruction score, the \textbf{anomaly score}. After evaluating the full test dataset, the resulting distribution of anomaly scores (see Figure~\ref{fig:perfectHisto}) can be used to define a threshold for anomaly detection. In a realistic scenario, such a threshold could be derived from background simulation and subsequently applied to data to flag events that (even slightly) deviate from the learned background structure, or could be established first in control regions before being used in signal regions. The anomaly score can directly replace the signal-region-defining variable in a standard LHC search: background estimation and statistical analysis proceed exactly as in a supervised analysis, with the supervised classifier output simply replaced by the anomaly score \cite{Aarrestad_2022}.

%%%%%%%%%%%%%%%%%%%%%%%%%%%%%%%%%
\subsection{VQ-VAE} 
\label{sec:VQVAE}
%%%%%%%%%%%%%%%%%%%%%%%%%%%%%%%%%%

To obtain a learned discrete representation of continuous collider event features, we employ a VQ-VAE tokenizer~\cite{oord2018} in the spirit of Ref.~\cite{Birk_2024}. This allows continuous event information to be compressed into discrete token sequences suitable for the downstream masked-token anomaly detection model.

Each event is represented as a padded sequence of reconstructed objects, where each object is described by four kinematic features, ($E, p_T,\eta,\phi$), and an associated binary mask indicating valid (non-padded) entries. 
One additional event-level token, carrying the missing transverse energy information (its magnitude and azimuthal angle $\phi$), is prepended and encoded in the same four-feature format.
These continuous four-feature tokens are not discrete yet: they are first embedded by the encoder into latent vectors, and only then quantized by the VQ codebook to integer token IDs used by the downstream BERT-like model.

The tokenizer is a masked sequence-to-sequence VQ-VAE with a NormFormer backbone \cite{shleifer2021normformer}.
For a mini-batch of events, the input tensor is
\[
x\in\mathbb{R}^{B\times S\times F},
\]
where \(B\) is the batch size, \(S\) the sequence length (number of tokens per event, including special tokens), and \(F\) the feature dimension per token (\(F=4\) in our setup).
The encoder applies a linear projection \(F\!\rightarrow\!H\) followed by \(N\) NormFormer blocks (multi-head self-attention + MLP with residual connections and LayerNorm),
and finally a latent projection \(H\!\rightarrow\!D\), producing continuous latent embeddings \[z_e\in\mathbb{R}^{B\times S\times D},\] with \(H\) the hidden width and \(D\) the latent dimension.

A vector-quantization layer with codebook
\[
C=\{c_k\}_{k=1}^{K},\qquad c_k\in\mathbb{R}^{D},
\]
discretizes each token position independently:
\[
k_{b,s}=\arg\min_{k\in\{1,\dots,K\}}\|z_{e,b,s}-c_k\|_2,\qquad
z_{q,b,s}=c_{k_{b,s}}.
\]
Here, \(K\) is the vocabulary size (number of codebook entries), \(k_{b,s}\in\{1,\dots,K\}\) is the discrete token ID at batch index \(b\), position \(s\), and \(z_q\in\mathbb{R}^{B\times S\times D}\) are the quantized latents passed to the decoder.

Thus, each event is mapped to an integer sequence \(\{k_{b,s}\}_{s=1}^{S}\) over a vocabulary of size \(K\) (including the prepended MET token).

The training objective is
\[
\mathcal{L}
=
\mathcal{L}_{\mathrm{reco}}
+
\alpha\,\mathcal{L}_{\mathrm{VQ}}
+
\beta\,D_{\mathrm{KL}}\!\left(p_{\mathrm{use}}\|U_K\right),
\]
where $\mathcal{L}_{\mathrm{reco}}$ is the MSE reconstruction loss, and 
$\mathcal{L}_{\mathrm{VQ}}$ is the standard VQ-VAE loss \cite{oord2018}, consisting of a codebook term and a commitment term,
\begin{equation}
\mathcal{L}_{\mathrm{VQ}}
=
\left\| \mathrm{sg}[z_e] - z_q \right\|_2^2
+
\gamma \left\| z_e - \mathrm{sg}[z_q] \right\|_2^2,
\end{equation}
where $\mathrm{sg}[\cdot]$ denotes the stop-gradient operator and $\gamma$ controls the strength of the commitment term. 
The first term updates the codebook embeddings, while the second encourages the encoder outputs to remain close to the selected codebook vectors. 
Finally, $p_{\mathrm{use}}$ is the empirical batch-wise code usage distribution over $K$ entries (with $U_K$ the uniform distribution). 
The use of the KL term encourages uniform codebook utilization, mitigating code collapse and improving token diversity \cite{Zhang_2023_CVPR}.

Hyperparameters are optimized with Optuna by minimizing validation loss.
To ensure controlled comparisons, the vocabulary size \(K\) is fixed within each optimization campaign, and all other hyperparameters are tuned conditionally on that fixed \(K\).
We repeat this procedure for multiple vocabulary sizes, running an independent Optuna study for each \(K\). In particular, \(\alpha\) and \(\beta\) are optimized jointly with architecture/training hyperparameters.

For each trial, we train with RAdam and a ReduceLROnPlateau scheduler; trials with unstable validation behavior (NaN/Inf) are pruned.

Because the downstream task is unsupervised anomaly detection, the tokenizer is trained and validated using background-only events. After training, the VQ-VAE weights are frozen, and the pretrained encoder+quantizer are applied to all processes (background and signal) to produce discrete token sequences for the downstream BERT-like masked-token anomaly detector.

For reproducibility, we provide the best-performing configurations of our model (for each fixed codebook size $K$) in Appendix~\ref{app:HyperparameterVQVAE}.

%%%%%%%%%%%%%%%%%%%%%%%%%%%%%%%%%%%%%%%%%%%%%%%%%%%%%
\section{Tokenization} 
\label{sec:Tokenization}
%%%%%%%%%%%%%%%%%%%%%%%%%%%%%%%%%%%%%%%%%%%%%%%%%%%%%%

An effective tokenization strategy is crucial to enable the masked-token prediction model to capture the underlying structure of background events during training. In this work, we investigate several approaches, including deep-learning-based methods such as VQ-VAE and discretization-based methods based on a LUT. We also study which types of event information, such as particle types, charges, kinematic variables, and global event quantities, should be encoded in the event sequence to optimize downstream performance. The following subsections describe and compare these tokenization strategies.

%To ensure that the LLM-like model accurately captures the underlying structure of background events during training, selecting an effective tokenization strategy is crucial. Therefore, several approaches were investigated in this work, including deep learning-based techniques (e.g. VQ-VAE) and discretization-based methods, such as a look-up table (LUT). Additionally, the types of event information (particle types, charges, kinematic variables, global event quantities), that should be encoded within the sequence of an event, to optimize the model performance were examined. The following subsections detail and compare these tokenization strategies.

%%%%%%%%%%%%%%%%%%%%%%%%%%%%%%%%%
\subsection{Tokenization via look-up table} 
\label{sec:binning}
%%%%%%%%%%%%%%%%%%%%%%%%%%%%%%%%%%

To represent events as discrete tokens, a LUT tokenization scheme was implemented, with careful considerations for both the number of bins and the definition of bin edges. Each token encodes either (i) a reconstructed particle object with its kinematics ({particle type and electric charge,  $p_T$, $\eta$, $\phi$}), or (ii) global event information, namely the missing transverse energy or its azimuthal angle. 

%Continuous variables are discretized into $K$ bins using quantile binning based on background events only, such that each bin contains approximately the same number of background entries. The particle type and charge, categorical attribute, keeps the inherited 7 classes (see Table~\ref{table:particle-object}). Multiples choices for K and for the bin-edge definitions were explored. $\phi$ and $\phi_{E^{miss}_T}$, periodic variables, are binned on $[-\pi,\pi)$. Binning is performed on the absolute pseudorapidity $\eta$. $p_T$ and $\|E^{miss}_T\|$ are discretized on $[0,+\infty)$. Tokens are emitted following the fixed order of the input event sequence (by object type, then decreasing transverse momentum, see Section~\ref{sec:DMdata}). Padding entries, inherited from the original dataset, are represented by a dedicated token (index 0).

The variables $p_T$, $|\eta|$, and $\|E^{\mathrm{miss}}_T\|$ are discretized into $N$ bins using quantile binning based on background events only, such that each bin contains approximately the same number of background entries. The particle type and charge are treated as categorical attributes and retain the seven classes inherited from the original dataset, see Table~\ref{table:particle-object}. The periodic variables $\phi$ and $\phi_{E^{\mathrm{miss}}_T}$ are instead binned uniformly on $[-\pi,\pi)$ into $N$ equal-width bins. Multiple choices of $N$ and of the bin-edge definitions were explored. 

Tokens are emitted following the fixed order of the input event sequence, first by object type and then by decreasing transverse momentum. Padding entries, inherited from the original dataset, are represented by a dedicated token with index 0.

Each tokenization configuration is evaluated downstream, by training the LLM-inspired model across various hyperparameters and measuring the separation power of the anomaly score (as described in Section~\ref{sec:LLManomalydetector}) between background and signal. Including in the event-sequence, both a $\|E^{miss}_T\|$ and a $\phi_{E^{miss}_T}$ token consistently improves downstream performance. All reported results therefore include them. The best-performing configurations are summarized below, while the detailed bin definitions are reported in Appendix~\ref{app:Bintokens}.

\noindent In the best-performing configurations:
\begin{itemize}
    \item[-]$p_T$, $\eta$ and $\|E^{miss}_T\|$ were each divided into $N$ bins ($N=[4,6]$), with equal occupancy.
    \item[-]$\phi$ and $\phi_{E^{miss}_T}$ were divided into $N$ bins of width $\frac2N\pi$.
    \item[-]Bins are indexed from 1 to $N$ and defined as half‑open intervals with the lower boundary included.
    \item[-]Each particle token is defined as follows: \\
    \(\texttt{token}_{part}=N^3\times(\text{bin}_{obj}-1)+N^2\times(\text{bin}_{p_T}-1)+N\times(\text{bin}_{\eta}-1)+\text{bin}_{\phi}\).
    \item[-]It yields: \hspace{1cm} 
    \begin{itemize*}[leftmargin=*,itemjoin={\quad}]
        \item[$\cdot$] \(\texttt{token}_{part}\in[1,7\times N^3]\) or \(\texttt{token}_{part}=0\) if padding;
    \end{itemize*}\\ 
    \begin{itemize*}[itemjoin=\quad, leftmargin=5cm, before=\null, after=\hskip1.9em\hfill]
        \item[$\cdot$]\(\texttt{token}_{\|E^{miss}_T\|}\in[7\times N^3+1,7\times N^3+N]\); \hspace{1.3cm} 
        \item[$\cdot$]\(\texttt{token}_{\phi_{E^{miss}_T}}\in[7\times N^3+N+1,7\times N^3+2N]\).
    \end{itemize*}
    \item[-]An event is represented as a sequence:\\
\([\texttt{token}_{part,1}, \texttt{token}_{part,2}, \texttt{token}_{part,3}, ..., \texttt{token}_{part,18}, \texttt{token}_{\|E^{miss}_T\|}, \texttt{token}_{\phi_{E^{miss}_T}}]\)
\end{itemize}

The discretizations of the best-performing configurations are reported in Appendix~\ref{app:Bintokens}, Tables~\ref{tab:binK4}, \ref{tab:binK5}, and~\ref{tab:binK6}. The token-index distributions for signal and background are shown in Appendix~\ref{app:Visualisation}, Figures~\ref{fig:EncodingBin} and Figure~\ref{fig:EncodingBin-BSM} for the four-top and the SUSY $\tilde{g}\tilde{g}$ scenario, respectively.

%%%%%%%%%%%%%%%%%%%%%%%%%%%%%%%%%
\subsection{Tokenization via VQ-VAE} 
\label{sec:tokenVQVAE}

As a second approach, we consider a learned tokenization based on the VQ-VAE described in Section~\ref{sec:VQVAE}. The resulting tokenized sequence has the structure
\[
[\texttt{token}_0,\texttt{token}_{\mathrm{part},1},\texttt{token}_{\mathrm{part},2},\texttt{token}_{\mathrm{part},3},\ldots,\texttt{token}_{\mathrm{part},18}],
\]
where $\texttt{token}_0$ encodes $\lVert E_T^{\mathrm{miss}} \rVert$ and $\phi_{E_T^{\mathrm{miss}}}$, and $\texttt{token}_{\mathrm{part},i}=0$ for padded entries.

To study the effect of the codebook size on both the VQ-VAE reconstruction performance and the anomaly detection power of the downstream model, we performed a scan over several codebook sizes. The most effective tokenization strategies were obtained for codebook sizes of 512, 850, and 1700, which appear to provide a favorable balance between representational compactness and downstream separation performance. The corresponding token-index distributions are shown in Appendix~\ref{app:Visualisation}, Figure~\ref{fig:EncodingVQ} for the four-top benchmark and Figure~\ref{fig:EncodingVQ-BSM} for the SUSY \(\tilde g \tilde g\) benchmark.

%%%%%%%%%%%%%%%%%%%
\section{Results and Evaluation}
\label{sec:Results}
%%%%%%%%%%%%%%%%%%%%

%%%%%%%%%%%%%%%%%%%
%\subsection{Comparison of Tokenization Strategies}
\subsection{Experimental setup for tokenization comparison}
\label{sec:besttokenization}
%%%%%%%%%%%%%%%%%%%%

%The best-performing VQ-VAE and LUT tokenization schemes are compared for both the four-top and BSM benchmarks. Most of the core architecture of the downstream model was kept fixed across all evaluations, with the following settings. The embedding dimension is 64, the number of transformer layers is 2 with 4 self-attention heads.  However, for each strategy, different scans of hyperparameters (covering dropout rate, type of positional encodings or its omission, learning rate of the ADAM optimizer, expansion factor of the feed-forward network…) were performed to ensure a fair and optimal evaluation. 
%% The specific model configurations used during the training that yields the best downstream performance for each tokenization schemes are summarized in Appendix~\ref{app:HyperparametersLLM}.  
%The specific model configurations used in the training that yield the best downstream performance for each tokenization scheme are summarized in Appendix~\ref{app:HyperparametersLLM}, in Tables~\ref{tab:downstream_hyperparameters} and~\ref{table:hyperparametersTokenization}.

The best-performing LUT and VQ-VAE tokenization schemes are compared for both the four-top and BSM benchmarks. To ensure a fair comparison, most of the core architecture of the downstream transformer-based model is kept fixed across all evaluations. In particular, the embedding dimension is 64, and the model consists of two transformer layers with four self-attention heads each.

For each tokenization strategy, dedicated hyperparameter scans are nevertheless performed to optimize downstream performance. These scans include the dropout rate, the choice of positional encoding (or its omission), the Adam learning rate, and the expansion factor of the feed-forward network, among other settings. The specific configurations yielding the best downstream performance for each tokenization scheme are summarized in Appendix~\ref{app:HyperparametersLLM}, in Tables~\ref{tab:downstream_hyperparameters_4top} and \ref{tab:downstream_hyperparameters_BSM}, for the four-top and BSM benchmarks, respectively.

The input vocabulary size of the downstream model is determined by the tokenization strategy itself: by the number of bins in the LUT approach and by the codebook size in the VQ-VAE approach.

The training of the masked-token prediction model itself is performed on a single standard GPU hardware and inference can be realized on CPU.

%%%%%%%%%%%%%%%%%%%%%%%%%%%%%%%%%%%%
\subsection{Performance comparison}
\label{sec:performance}
%%%%%%%%%%%%%%%%%%%%%%%%%%%%%%%%%%%%

Performance is assessed using the area under the receiver operating characteristic (ROC) curve, quantified by the area under the curve (AUC), computed after the inference procedure described in Section~\ref{sec:LLManomalydetector}.
For a given tokenization scheme, the downstream model is trained exclusively on background events encoded according to that scheme. During inference, both signal and background events, encoded with the same scheme, are reconstructed by the trained model and assigned anomaly scores, from which the AUC is computed. The yielded distribution of anomaly scores can be found in Appendix~\ref{app:Histograms}

\begin{figure}[h!]
\begin{subfigure}{0.49\textwidth}
\includegraphics[width=\linewidth]{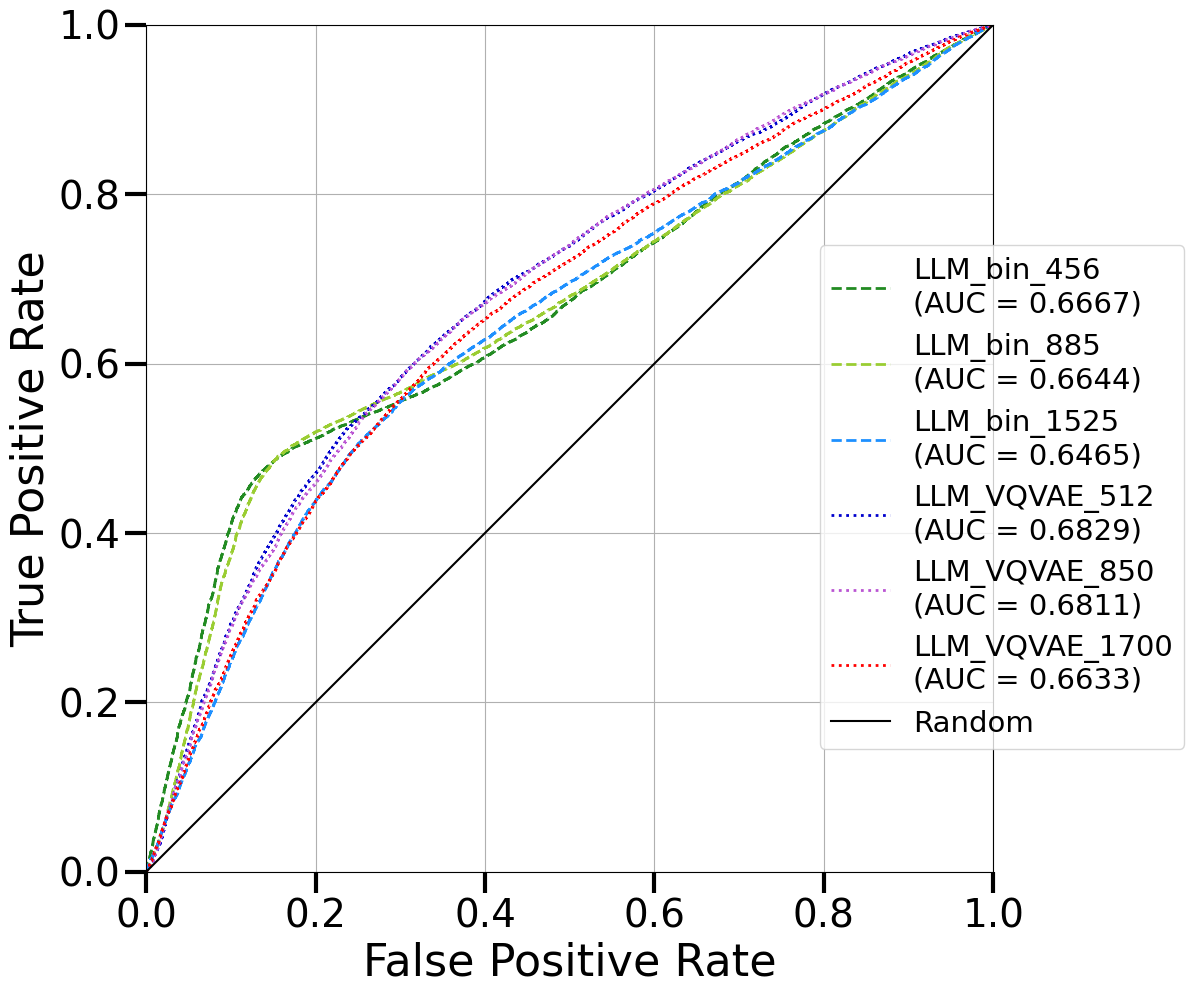} 
\caption{$t\bar{t}t\bar{t}$ scenario.}
\label{fig:comparisonTokenization}
\end{subfigure}
\hfill
\begin{subfigure}{0.49\textwidth}
\includegraphics[width=\linewidth]{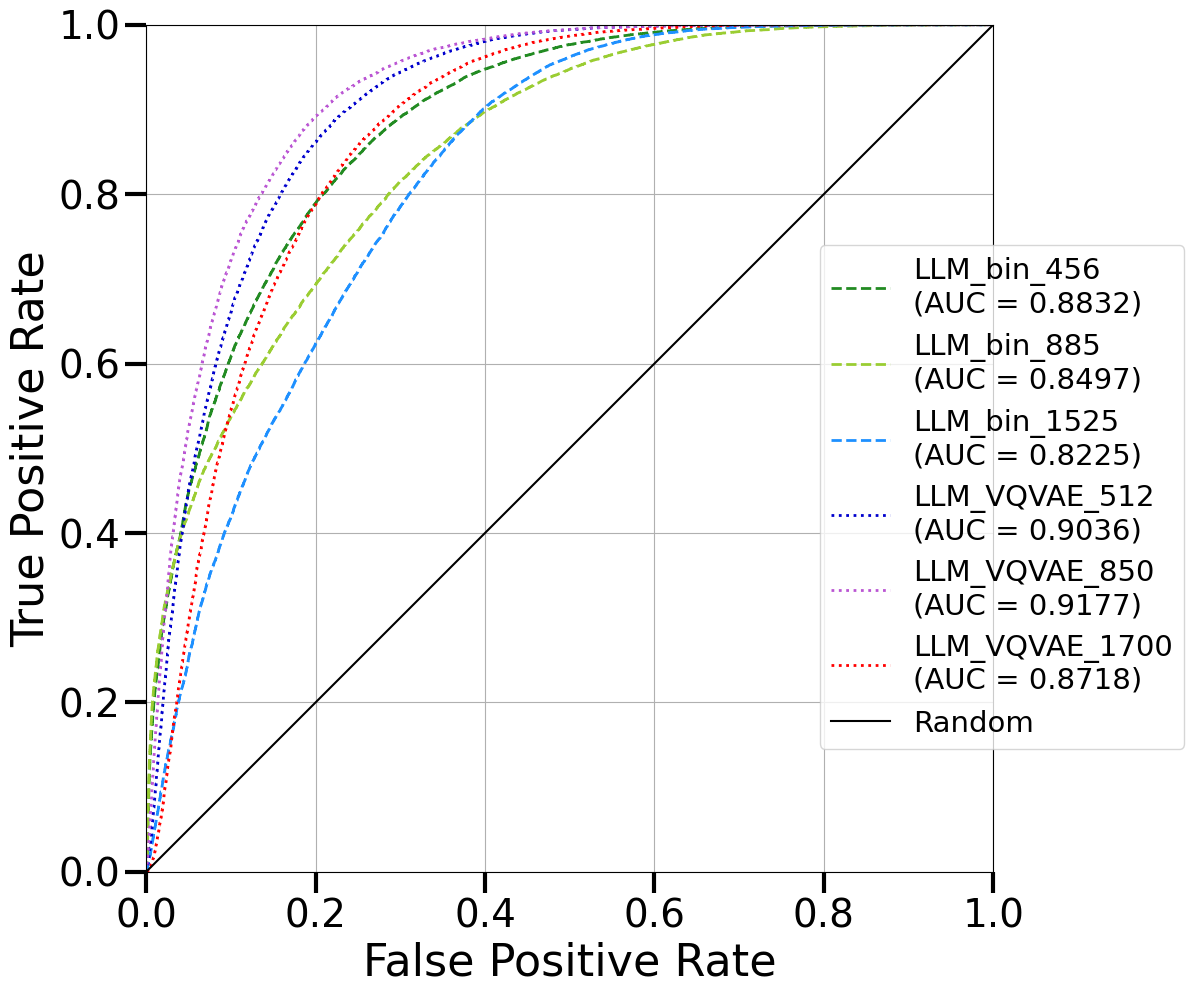} 
\caption{$\tilde{g}\tilde{g}$ scenario.}
\label{fig:comparisonTokenization-BSM}
\end{subfigure}
\caption{ROC curves for the downstream models evaluated with each tokenization strategy, where LUT denotes the look-up table tokenization, shown for the $t\bar{t}t\bar{t}$ scenario on the left and the $\tilde{g}\tilde{g}$ scenario on the right.}
\end{figure}

Results for the four-top benchmark are presented in Table~\ref{table:comparisonTokenization} and in Figure~\ref{fig:comparisonTokenization}, while those for the BSM benchmark are shown in Table~\ref{table:comparisonTokenization-BSM} and in Figure~\ref{fig:comparisonTokenization-BSM}. In both benchmarks, the best-performing configuration is obtained with the VQ-VAE tokenization, although the size of the improvement depends strongly on the physics scenario. In the four-top benchmark, where signal and background topologies are highly similar, the gain over the best LUT configuration is modest: the ROC AUC increases from 0.6667 to 0.6829. This limited separation is also visible in Figure~\ref{fig:comparisonTokenization}, where all curves remain relatively close to one another. 

\begin{table}[t] %[htbp]
% \centering
% \begin{table}[htb!]
\caption{Performance of the downstream models for each tokenization strategy in terms of AUC, signal efficiency ($\epsilon_S$) evaluated at background efficiency ($\epsilon_B = 0.01$), and background efficiency ($\epsilon_B$) evaluated at signal efficiency ($\epsilon_S = 0.01$).}
\small
\setlength{\tabcolsep}{5pt}
\renewcommand{\arraystretch}{1.15}
\begin{subtable}[t]{0.49\textwidth}
\centering
\caption{$t\bar{t}t\bar{t}$ scenario.}
\scalebox{0.73}{
\begin{tabular}{ccccc}
\toprule
\textbf{Token. scheme} & \textbf{Voc. size} & \textbf{AUC} & \textbf{$\epsilon_B(\epsilon_S=0.01)$} & \textbf{$\epsilon_S(\epsilon_B=0.01)$} \\
\midrule
\multirow{3}{*}{LUT}
  & 456  & 0.6667 & 0.002 & 0.043 \\
  & 885  & 0.6644 & 0.005 & 0.024 \\
  & 1524 & 0.6465 & 0.004 & 0.024 \\
\addlinespace
\midrule
\addlinespace
\multirow{3}{*}{VQ-VAE}
  & 512  & 0.6829 & 0.007 & 0.19  \\
  & 850  & 0.6811 & 0.007 & 0.19  \\
  & 1700 & 0.6633 & 0.005 & 0.024 \\
\bottomrule
\end{tabular}
}
\label{table:comparisonTokenization}
\end{subtable}
\hfill
\setlength{\tabcolsep}{5pt}
\renewcommand{\arraystretch}{1.15}
\begin{subtable}[t]{0.49\textwidth}
\centering
\caption{$\tilde{g}\tilde{g}$ scenario.}
\scalebox{0.73}{
\begin{tabular}{ccccc}
\toprule
\textbf{Token. scheme} & \textbf{Voc. size} & \textbf{AUC} & \textbf{$\epsilon_B(\epsilon_S=0.01)$} & \textbf{$\epsilon_S(\epsilon_B=0.01)$} \\
\midrule
\multirow{3}{*}{LUT}
  & 456  & 0.8832 & 0.001 & 0.211 \\
  & 885  & 0.8497 & 0.001 & 0.227 \\
  & 1524 & 0.8225 & 0.002 & 0.051 \\
\addlinespace
\midrule
\addlinespace
\multirow{3}{*}{VQ-VAE}
  & 512  & 0.9036 & 0.002 & 0.087 \\
  & 850  & 0.9177 & 0.001 & 0.149 \\
  & 1700 & 0.8718 & 0.007 & 0.023 \\
\bottomrule
\end{tabular}
}

\label{table:comparisonTokenization-BSM}
\end{subtable}
\end{table}
%%%%%%

For the LUT scheme, the ROC curves obtained in the four-top benchmark exhibit non‑monotonic behaviour, indicating partially overlapping score distributions. While this feature becomes less pronounced at larger vocabulary sizes, it persists even at 1524 %\PM{or 1525?}, %\PM{and the histograms in Figures~\ref{fig:Histogram456}, \ref{fig:Histogram885}, and \ref{fig:Histogram1525} in Appendix~\ref{app:Histograms} confirm} 
that a fraction of signal and background events remain indistinguishable regardless of vocabulary size (see in Appendix~\ref{app:Histograms}, Figures \ref{fig:Histogram456}, \ref{fig:Histogram885} and \ref{fig:Histogram1524} for the vocabulary sizes $456$, $885$ and $1524$, respectively). 
We attribute this to the intrinsic similarity between signal and background in the four-top case. With an insufficient vocabulary, more signal events are encoded identically to background events, capping the attainable discrimination. Increasing the vocabulary size yields a smoother ROC curve which does not translate into an improvement in overall AUC.

% \PM{I think the vocabulary sizes for LUT and VQ-VAE may have been mixed up, so the corresponding references and descriptions should be checked and updated carefully. Also, the references to the SUSY histograms are missing. \\
% For \texttt{tttt}:\\
% LUT: \ref{fig:Histogram456}, \ref{fig:Histogram885}, \ref{fig:Histogram1524};
% VQ-VAE: \ref{fig:Histogram512}, \ref{fig:Histogram850}, \ref{fig:Histogram1700}.\\
% For SUSY:\\
% LUT: \ref{fig:Histogram456-BSM}, \ref{fig:Histogram885-BSM}, \ref{fig:Histogram1524-BSM};
% VQ-VAE: \ref{fig:Histogram512-BSM}, \ref{fig:Histogram850-BSM}, \ref{fig:Histogram1700-BSM}.}\\
%\AV{should be corrected now.}\\
The dependence on vocabulary size is non-trivial. For the VQ-VAE, increasing the codebook size from 512 to 850 leaves the performance in the four-top benchmark essentially unchanged, but noticeably improves the BSM benchmark. This suggests that a finer learned discretization is beneficial when the signal departs more strongly from the background manifold. However, increasing the vocabulary further to 1700 degrades performance in both benchmarks. A similar deterioration is observed for the largest LUT vocabularies. This indicates that an overly fine discretization can fragment the representation, reduce token statistics, and make it harder for the downstream model to learn stable background patterns.

Overall, these results indicate that learned tokenization is advantageous in this setup, particularly for signals with more distinctive event structure, while the more challenging four-top case remains limited by the intrinsic similarity between signal and background rather than by tokenization alone.

%\RR{We should explain the weird shape of the ROC for the 4tops and bining case}\AV{done}

%Among the evaluated approaches, the highest ROC AUC is achieved with the VQ-VAE tokenization for both benchmarks, indicating that the deep learned scheme provides a more faithful discretization of the event information and enables more effective separation of signal deviations from the learned background. The choice of vocabulary size (between 512 and 850) in the VQ-VAE has a more pronounced impact on the BSM benchmark than on the SM case. For the SUSY $\tilde{g}\tilde{g}$ study, increasing the codebook size from 512 to 850 leads to more than a 1\% improvement in ROC AUC, whereas no comparable gain is observed for the $t\bar{t}t\bar{t}$ scenario, where the signal–background differences are inherently subtle. Overall, rich token granularity benefits the more distinct BSM topology, but only up to a point; beyond that, the token space seems to become too sparse to support effective anomaly discrimination. Further increasing the vocabulary size to 1700 leads in fact to a significant deterioration in performance for both benchmarks, indicating that excessively large codebooks fragment the representation and hinder the model’s ability to learn the background distribution or facilitate its signal reconstruction. 

%%%%%%%%%%%%%%%%%%%
\subsection{Comparison with Established Unsupervised Methods}
\label{sec:comparison_unsup}
%%%%%%%%%%%%%%%%%%%%%

In high-energy physics, a broad landscape of anomaly detection methods has been developed around three axes: the supervision level, the signal topology and the (algorithmic) strategy; and a key distinction exists between two complementary paradigms. The first, exemplified by the LHC Olympics challenge (Ref.~\cite{Kasieczka:2021xcg}) targets resonance searches. 
%it aims at discovering a localised excess in an invariant mass distribution
The anomaly detection task is to enhance the bump-hunt sensitivity using sideband-based methods. The second strategy, which is followed here, targets the definition of model-independent anomaly scores that can replace the signal-region-defining variable in a standard LHC search.
% once an anomaly score is defined, the background estimation and statistical analysis proceed exactly as in a supervised analysis, with the supervised classifier output simply replaced by the anomaly score
This approach, pioneered by the Dark Machines Challenge \cite{Aarrestad_2022}, is topology-agnostic and requires no resonant variable or sideband structure. 
The challenge and its resulting paper (Ref.~\cite{Aarrestad_2022}) showed that performances of the different methods depend greatly on the signal topologies. We therefore chose to compare our work with methods applicable in the same setting that yield great performance on our selected benchmarks: the DeepSVDD and the DROCC method, one-class classifiers based respectively on hypersphere minimization and adversarial binary classification, and the DDD method, a weakly supervised classifier converted for anomaly detection through a strategy of artificial distortions of the background. These are well-established unsupervised methods from Ref.~\cite{Moskvitina:2025epjc}.
\subsubsection{Four-top benchmark}
\label{sec:results4tops}
%%%%%%%%%%%%%%%%%%%%%%

Separating the $t\bar{t}t\bar{t}$ signal, with its complex multi-jet topology, from the background ($t\bar{t}W$, $t\bar{t}WW$, $t\bar{t}Z$, $t\bar{t}H$) is a particularly challenging case for anomaly detection. %\AV{\sout{We therefore compare our results with established unsupervised methods from Ref.~\cite{Moskvitina:2025epjc} tested on the same benchmark}}. 
Figure~\ref{fig:comparisonPolina} shows the corresponding ROC curves and Table~\ref{table:comparisonPolinas}, the evaluation metrics. AUC values range from 0.48 to 0.75, reflecting the well-known kinematic similarity between the four-top signal and the SM background. Several published baselines perform at or below random in this regime, highlighting the intrinsic difficulty of this topology. The masked-token prediction approach achieves an AUC of 0.683, outperforming all DeepSVDD and DROCC baselines and ranking second only to the DDD variants. This is particularly noteworthy given the discrete nature of the tokenized representation, suggesting that the codebook-based encoding captures event-level features that remain robust despite the signal-background similarity characteristic of the four-top topology.

%%%%%%%%%%%%%%%%%%%%%%%%
\subsubsection{BSM benchmark}
\label{sec:resultsBSM}
%%%%%%%%%%%%%%%%%%%%%%

\begin{figure}[h!]
  % ---- FIGURE  ----
    \begin{subfigure}{0.49\textwidth}
      \includegraphics[width=\linewidth]{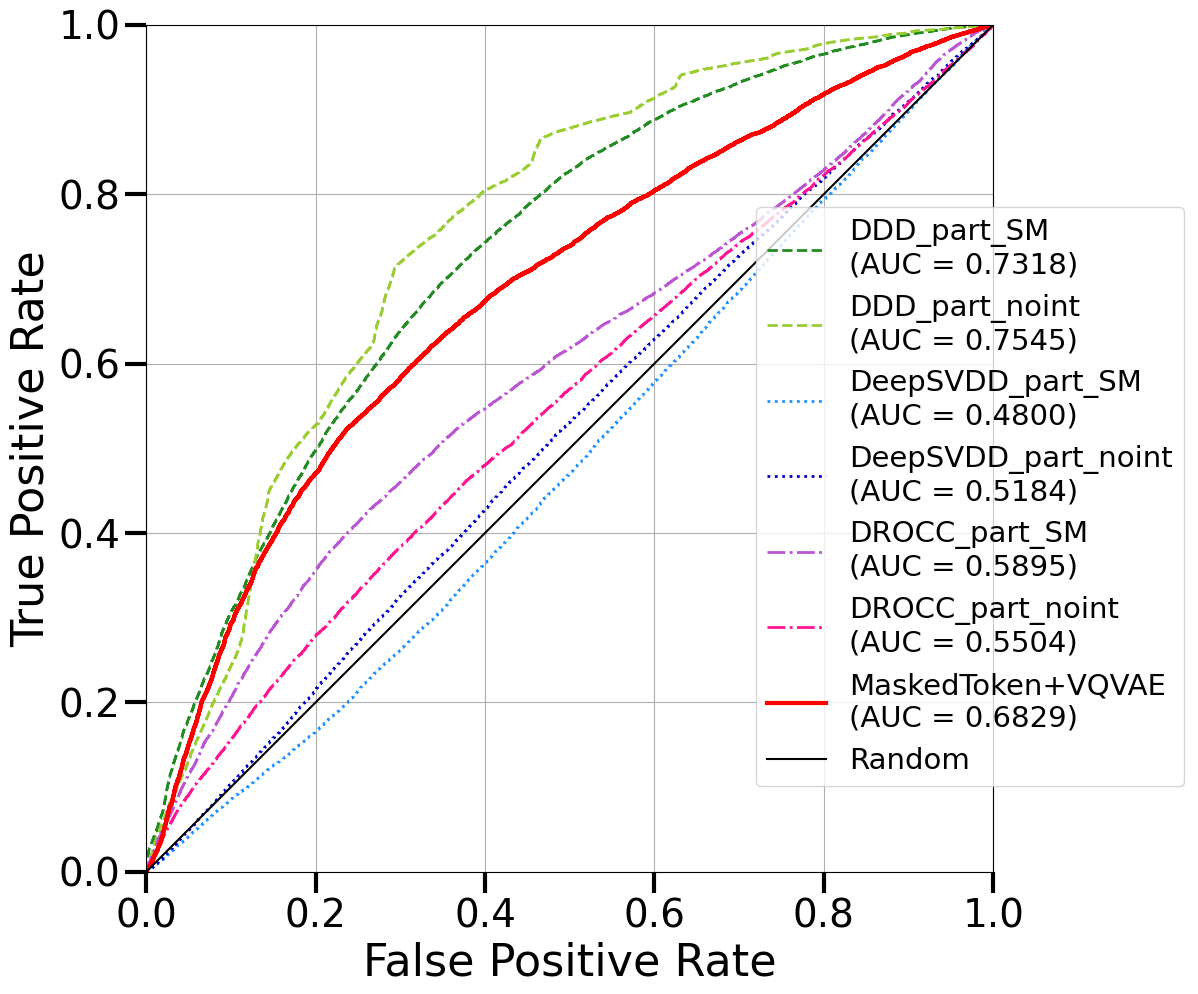}
      \caption{$t\bar{t}t\bar{t}$ scenario.}
      \label{fig:comparisonPolina}
    \end{subfigure}
    \hfill
    \begin{subfigure}{0.49\textwidth}
      \includegraphics[width=\linewidth]{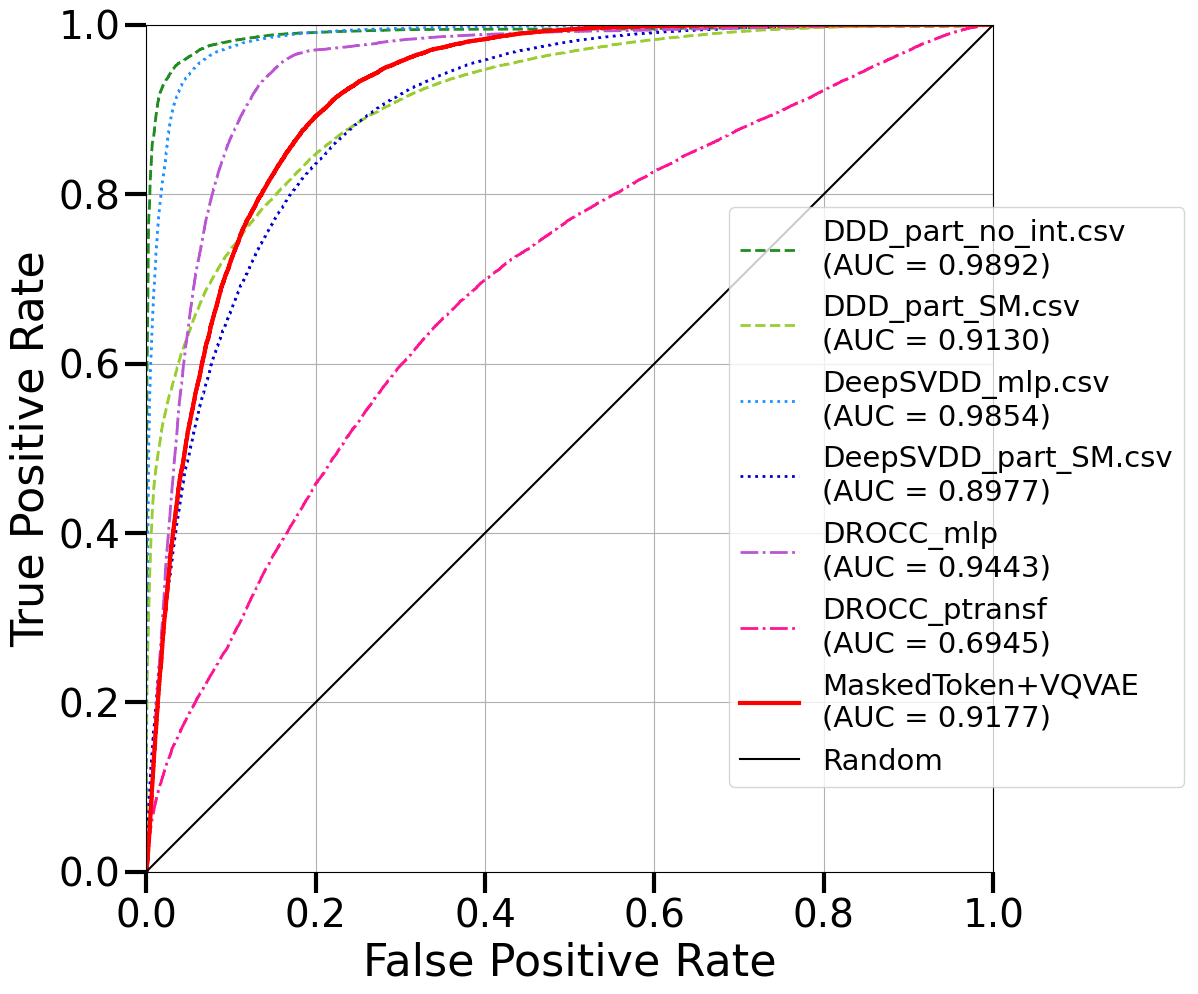}
      \caption{$\tilde{g}\tilde{g}$ scenario.}
      \label{fig:comparisonPolina-BSM}
    \end{subfigure}
    \caption{ROC curves for the proposed method (labelled as `MaskedToken+VQVAE' in the
      legend) and other established unsupervised methods from Ref.~\cite{Moskvitina:2025epjc},
      shown for the $t\bar{t}t\bar{t}$ scenario on the left and the $\tilde{g}\tilde{g}$
      scenario on the right.}
\end{figure}

The separation of SUSY gluino-gluino pair production from the SM background is an easier discrimination task. Figure~\ref{fig:comparisonPolina-BSM} shows the ROC curves of the masked-token prediction method together with those of the unsupervised methods presented in Ref.~\cite{Moskvitina:2025epjc} that were tested on the gluino-pair production benchmark, and Table~\ref{table:comparisonPolinas-BSM}, the evaluation metrics. AUC values across all shown methods are high, ranging from 0.69 to 0.99, which is consistent with good kinematic separation. Our approach achieves an AUC of 0.918, placing it competitively among the published baselines tested on the same dataset. This demonstrates that a learned encoding coupled with masked-token prediction retains sufficient kinematic information for anomaly detection, despite operating on a discrete tokenized representation of the event rather than continuous inputs.

\begin{table}[t] %[htbp]
  % ---- TABLE  ----
\caption{For comparison with our masked-token prediction model, performance of the models of Ref.~\cite{Moskvitina:2025epjc} in terms of AUC, signal efficiency ($\epsilon_S$) evaluated at background efficiency ($\epsilon_B = 0.01$), and background efficiency ($\epsilon_B$) evaluated at signal efficiency ($\epsilon_S = 0.01$).}
\small
\setlength{\tabcolsep}{5pt}
\renewcommand{\arraystretch}{1.15}
\begin{subtable}[t]{0.49\textwidth}
\centering
\caption{$t\bar{t}t\bar{t}$ scenario.}
\scalebox{0.73}{
\begin{tabular}{lcccc}
\toprule
\textbf{Model name} & \textbf{AUC} & \textbf{$\epsilon_B(\epsilon_S=0.01)$} & \textbf{$\epsilon_S(\epsilon_B=0.01)$} \\
\midrule
          MaskedToken + VQVAE  & 0.6829 & 0.007 & 0.19  \\
          DDD\_part\_SM        & 0.7318 & 0.001 & 0.043 \\
          DDD\_part\_noint     & 0.7545 & 0.003 & 0.033 \\
          DeepSVDD\_part\_SM   & 0.4800 & 0.014 & 0.007 \\
          DeepSVDD\_part\_noint& 0.5184 & 0.013 & 0.07  \\
          DROCC\_part\_SM      & 0.5895 & 0.004 & 0.026 \\
          DROCC\_part\_noint   & 0.5504 & 0.004 & 0.022 \\
          \bottomrule
        \end{tabular}
      }
      \label{table:comparisonPolinas}
    \end{subtable}
    \hfill
    \begin{subtable}[t]{0.49\textwidth}
      \centering
      \caption{$\tilde{g}\tilde{g}$ scenario.}
      \scalebox{0.73}{
        \begin{tabular}{lcccc}
          \toprule
          \textbf{Model name} & \textbf{AUC} & \textbf{$\epsilon_B(\epsilon_S=0.01)$} & \textbf{$\epsilon_S(\epsilon_B=0.01)$} \\
          \midrule
          MaskedToken + VQVAE  & 0.9177 & 0.001           & 0.149 \\
          DDD\_part\_SM        & 0.9130 & \textless 0.001 & 0.464 \\
          DDD\_part\_noint     & 0.9892 & \textless 0.001 & 0.880 \\
          DeepSVDD\_mlp        & 0.9854 & \textless 0.001 & 0.701 \\
          DeepSVDD\_part\_SM   & 0.8977 & 0.001           & 0.182 \\
          DROCC\_mlp           & 0.9443 & 0.002       & 0.160 \\
          DROCC\_ptransf       & 0.6945 & \textless 0.001 & 0.078 \\
          \bottomrule
        \end{tabular}
      }
      \label{table:comparisonPolinas-BSM}
    \end{subtable}
\end{table}
  \vspace{1em}

%%%%%%%%%%%%%%%%%%%%%%%%
\subsubsection{Position in the anomaly detection landscape}
\label{sec:position}
%%%%%%%%%%%%%%%%%%%%%%
Anomaly detection at the LHC is a notoriously difficult and diverse problem. Within the current extensive portfolio, no single method dominates across all signal topologies, therefore the field has converged on the view that a set of complementary approaches is needed \cite{Aarrestad_2022}. In this context, our method performs competitively across the tested benchmark, ranking only second to the DDD method for a four-top-quark signal. However a raw AUC comparison does not fully capture the benefits from the methods in terms of operating assumptions and broader applicability. While the DDD method achieves a higher AUC on both benchmarks, it requires a dedicated training procedure for each analysis scenario. This masked-token prediction method, by contrast, applies across different signal hypotheses without retraining. A single model pretrained on SM background can be directly used for anomaly detection on any signal topology. It is also architecturally compatible with foundation models as they are pretrained on token sequences with a masking objective, anomaly detection with our method becomes a natural byproduct of the foundation model training pipeline rather than a separate dedicated effort.

%%%%%%%%%%%%%%%%%%%%%%%%%%%%
\section{Conclusions}
\label{sec:Conclusions}
%%%%%%%%%%%%%%%%%%%%%%%%%%%%
In this work, we have explored the use of LLM-inspired anomaly detection strategies for high-energy physics events by formulating collider data as sequences of discrete tokens. We considered two different tokenization procedures, namely a deterministic LUT approach and a learned tokenization based on a VQ-VAE, and studied their impact on the performance of a masked-token prediction framework for unsupervised anomaly detection.

Our results show that the choice of tokenization plays an important role in the final anomaly detection performance. In both benchmark scenarios considered in this work, the VQ-VAE tokenization yielded the best overall results, although the size of the improvement depended on the physics case. For the four-top benchmark, where the signal is particularly challenging due to its close resemblance to the SM background, the gain with respect to the best LUT configuration is modest. By contrast, in the BSM benchmark the learned tokenization leads to a more visible improvement, indicating that it can preserve discriminating information more efficiently when the signal departs more strongly from the background manifold.

We also observed that the vocabulary size must be chosen with care. Increasing the number of tokens can improve the representation only up to a point, while excessively large vocabularies tend to degrade performance, likely because they fragment the data representation and reduce the statistical robustness of the downstream masked-token model. This behavior is seen for both tokenization strategies and highlights the existence of an optimal intermediate discretization scale.

Beyond the comparison between tokenization methods, our study shows that LLM-inspired masked-token approaches provide a viable and flexible framework for unsupervised searches at the LHC. The method is able to identify anomalous structures in challenging final states and performs competitively against more established unsupervised techniques. This supports the idea that generative and sequence-modeling tools originally developed for natural language processing can be successfully adapted to collider data analysis. In addition, looking forward, as foundation models for particle physics based on discrete token vocabularies become available, anomaly detection via masked-token prediction becomes a natural byproduct of their pretraining. No dedicated training is required, the anomaly score is obtained by comparing predicted and reconstructed tokens.

%\AV{Transferability is inherent to the method.}
%\CN{A model trained on SM background transfers across different BSM searches, thus enabling a model-independent anomaly detection method from a single training.} \AV{Scalability follows directly from this property. In most cases as the number of signal hypotheses to be tested grows, the computational cost does not, since no additional training is required per hypothesis. This is in particular the case for BSM signals versus a general standard model background or for SM signals that share the same background. Both of these benefits of the method also yield reduced computational cost, as no re-training is needed. The training itself is performed here on a single standard GPU hardware and is achieve in fewer than a hundred epochs.} 

%\AV{The choice of masked-token prediction compared to continuous masked-particle modeling also reduces the computational cost as the transformer architecture required is smaller. This choice also makes this method compatible with foundation models as they have masking-based pretraining objectives, meaning the anomaly score is a direct byproduct of the pretraining loss, requiring no additional training head nor a separate dedicated effort.}

Overall, our results indicate that learned tokenization, combined with masked-token anomaly detection, is a promising avenue for model independent new physics searches. %\AV{Together its properties position masked-token prediction as a complementary addition to the anomaly detection toolkit at the LHC.}

%%%%%%%%%%%%%%%%%%%%%%%%%%%%%%%
\section*{Acknowledgements}
%%%%%%%%%%%%%%%%%%%%%%%%%%%%%%%
The author(s) gratefully acknowledges the computer resources at Artemisa, funded by the European Union ERDF and Comunitat Valenciana as well as the technical support provided by the Instituto de Física Corpuscular, IFIC (CSIC-UV). The work of R. RdA was supported by PID2023-148162NB-C22 from the Spanish Ministerio de Ciencia e Innovación and by the PROMETEO/2022/69 from the Spanish GVA. 

\pagebreak 

\appendix

%%%%%%%%%%%%%%%%%%%%%%%%%%%%%%%
\section*{Appendix}
%%%%%%%%%%%%%%%%%%%%%%%%%%%%%%%

%%%%%%%%%%%%%%%%%%%%%%%%%%%%%%%
\section{Best-performing VQ-VAE configurations}
\label{app:HyperparameterVQVAE}
%%%%%%%%%%%%%%%%%%%%%%%%%%%%%%%

For each fixed codebook size $K$, we carried out an independent hyperparameter optimization of the VQ-VAE tokenizer. Tables~\ref{tab:VQVAE_best_4top} and~\ref{tab:VQVAE_best_SUSY} report the best-performing configurations obtained for the four-top and SUSY benchmarks, respectively. Each row corresponds to the optimal tokenizer setup found for a given codebook size and includes the main architectural and regularization hyperparameters used during training.

\setcounter{table}{0}
\renewcommand{\thetable}{A.\arabic{table}}
\begin{table}[htbp]
\centering
\caption{Best-performing VQ-VAE hyperparameter configurations for the four-top benchmark, for each reported codebook size $K$.}
\label{tab:VQVAE_best_4top}
\begin{tabular}{c c c c c c c}
\hline
$K$ & Latent dim & Hidden dim & Heads & Blocks & $\alpha$ & $\beta$ \\
\hline
512  & 10 & 64 & 2 & 2 & 0.4 & 0.1 \\
850  & 10 & 32 & 4 & 4 & 0.4 & 0.1 \\
1700 & 10 & 64 & 4 & 3 & 0.2 & 0.1 \\
\hline
\end{tabular}
\end{table}

\begin{table}[htbp]
\centering
\caption{Best-performing VQ-VAE hyperparameter configurations for the SUSY benchmark, for each reported codebook size $K$.}
\label{tab:VQVAE_best_SUSY}
\begin{tabular}{c c c c c c c}
\hline
$K$ & Latent dim & Hidden dim & Heads & Blocks & $\alpha$ &  $\beta$ \\
\hline
512  & 10 & 64  & 8 & 2 & 1.0 & 0.1 \\
850  & 5  & 128 & 8 & 2 & 2.0 & 0.1 \\
1700 & 10 & 128 & 4 & 4 & 0.5 & 0.16 \\
\hline
\end{tabular}
\end{table}

%%%%%%%%%%%%%%%%%%%%%%%%%%%%%%%
\section{Look-up tables for tokenization}
\label{app:Bintokens}
%%%%%%%%%%%%%%%%%%%%%%%%%%%%%%%

For each benchmark, the bin edges of each kinematic variable yielding the best-performing models are listed in Table~\ref{tab:binK4} for $N=4$,  Table~\ref{tab:binK5} for $N=5$, and  Table~\ref{tab:binK6} for $N=6$.

% \setcounter{table}{0}   
% \renewcommand{\thetable}{B.\arabic{table}}
% \begin{table}[h!]
% \caption{Bin definition for the Look-up table K=4.}
% \label{tab:binK4}
% \centering
% \begin{tabular}{c c c c c c c }
% \toprule
% \textbf{study} & \textbf{kinematic} & \textbf{bin edge}  & \textbf{bin edge} & \textbf{bin edge} & \textbf{bin edge} &\textbf{bin edge} \\ %\cline{1-7}
% \specialrule{0.75pt}{1pt}{1pt}
% \multirow{4}{*}{\textbf{$t\bar{t}t\bar{t}$}} & $p_T$ & $0$ & $10.47$ & $10.95$ & $11.41$ & $+\infty$\\ \cline{2-7}
%  & $\eta$ & $0$ & $0.4871$ & $1.009$ & $1.634$ & $+\infty$\\ \cline{2-7}
%  & $\phi$/$\phi_{E^{miss}_T}$ & $-\pi$ & $-\pi/2$& 0 &$\pi/2$ & $\pi$ \\ \cline{2-7}
%  & $\|E^{miss}_T\|$ & $0$ & $11.69$ & $12.22$ & $12.66$ & $+\infty$\\ %\cline{1-7} 
% \specialrule{0.75pt}{1pt}{1pt}
% \multirow{4}{*}{\textbf{SUSY $g\tilde{g}$}} & $p_T$ & $0$ & $10.75$ & $11.33$ & $12$ & $+\infty$\\ \cline{2-7}
%  & $\eta$ & $0$ & $0.4325$ & $0.9$ & $1.482$ & $+\infty$\\ \cline{2-7}
%  & $\phi$/$\phi_{E^{miss}_T}$ & $-\pi$ & $-\pi/2$& 0 &$\pi/2$ & $\pi$ \\ \cline{2-7}
%  & $\|E^{miss}_T\|$ & $0$ & $11.14$ & $11.42$ & $11.70$ & $+\infty$\\ 
% \bottomrule
% \end{tabular}
% \end{table}

\setcounter{table}{0}
\renewcommand{\thetable}{B.\arabic{table}}

\begin{table}[htb!]
\centering
\caption{Bin definitions for the look-up table with $N=4$. 
%\PM{It might be clearer to label the bin edge columns as `Bin edge 1', `Bin edge 2', `Bin edge 3', etc. (or simply `Bin 1', `Bin 2', `Bin 3', etc.), rather than repeating `Bin edge' in each column. The same comment applies to the other two tables.}
}
\label{tab:binK4}
\small
\setlength{\tabcolsep}{5pt}
\renewcommand{\arraystretch}{1.15}
\scalebox{0.9}{
\begin{tabular}{lcccccc}
\toprule
\textbf{Study} & \textbf{Kinematic variable} & \textbf{Bin edge 1} & \textbf{Bin edge 2} & \textbf{Bin edge 3} & \textbf{Bin edge 4} & \textbf{Bin edge 5} \\
\midrule
\multirow{4}{*}{$t\bar{t}t\bar{t}$}
& $p_T$                    & $0$     & $10.47$  & $10.95$ & $11.41$ & $+\infty$ \\
& $\eta$                   & $0$     & $0.4871$ & $1.009$ & $1.634$ & $+\infty$ \\
& $\phi$/$\phi_{E_T^{\mathrm{miss}}}$ & $-\pi$  & $-\pi/2$ & $0$    & $\pi/2$ & $\pi$ \\
& $\|E_T^{\mathrm{miss}}\|$            & $0$     & $11.69$  & $12.22$ & $12.66$ & $+\infty$ \\
\addlinespace
\midrule
\addlinespace
\multirow{4}{*}{SUSY $g\tilde{g}$}
& $p_T$                    & $0$     & $10.75$  & $11.33$ & $12$    & $+\infty$ \\
& $\eta$                   & $0$     & $0.4325$ & $0.9$   & $1.482$ & $+\infty$ \\
& $\phi$/$\phi_{E_T^{\mathrm{miss}}}$ & $-\pi$  & $-\pi/2$ & $0$    & $\pi/2$ & $\pi$ \\
& $\|E_T^{\mathrm{miss}}\|$            & $0$     & $11.14$  & $11.42$ & $11.70$ & $+\infty$ \\
\bottomrule
\end{tabular}
}
\end{table}

% \begin{table}[h!]
% \caption{Bin definition for the Look-up table K=5.}
% \label{tab:binK5}
% \centering
% \scalebox{0.999}{
% \centering
% \begin{tabular}{c c c c c c c c }
% \toprule
% \textbf{study} & \textbf{kinematic} & \textbf{bin edge}  & \textbf{bin edge} & \textbf{bin edge} & \textbf{bin edge} &\textbf{bin edge} &\textbf{bin edge}\\ %\cline{1-8}
% \specialrule{0.75pt}{1pt}{1pt}
% \multirow{4}{*}{\textbf{$t\bar{t}t\bar{t}$}} & $p_T$ & $0$ & $10.64$ & $11.10$ & $11.57$ & $12.17$ & $+\infty$\\ \cline{2-8}
%  & $\eta$ & $0$ & $0.3444$ & $0.7052$ & $1.109$ & $1.633$ & $+\infty$\\ \cline{2-8}
%  & $\phi$/$\phi_{E^{miss}_T}$ & $-\pi$ & $-3\pi/5$& $-\pi/5$ &$\pi/5$ & $3\pi/5$ & $\pi$ \\ \cline{2-8}
%  & $\|E^{miss}_T\|$ & $0$ & $11.55$ & $12.03$ & $12.40$ & $12.76$ & $+\infty$\\ %\cline{1-8}  
% \specialrule{0.75pt}{1pt}{1pt}
% \multirow{4}{*}{\textbf{SUSY $g\tilde{g}$}} & $p_T$ & $0$ & $10.37$ & $10.76$ & $11.13$ & $11.52$ & $+\infty$\\ \cline{2-8}
%  & $\eta$ & $0$ & $0.3877$ & $0.7918$ & $1.239$ & $1.789$ & $+\infty$\\ \cline{2-8}
%  & $\phi$/$\phi_{E^{miss}_T}$ & $-\pi$ & $-3\pi/5$& $-\pi/5$ &$\pi/5$ & $3\pi/5$ & $\pi$ \\ \cline{2-8}
%  & $\|E^{miss}_T\|$ & $0$ & $11.08$ & $11.31$ & $11.52$ & $11.78$ & $+\infty$\\ 
% \bottomrule
% \end{tabular}}
% \end{table}

\begin{table}[htb!]
\centering
\caption{Bin definitions for the look-up table with $N=5$.}
\label{tab:binK5}
\small
\setlength{\tabcolsep}{5pt}
\renewcommand{\arraystretch}{1.15}
\scalebox{0.88}{
\begin{tabular}{lccccccc}
\toprule
% \textbf{Study} & \textbf{Kinematic variable} & \textbf{Edge 1} & \textbf{Edge 2} & \textbf{Edge 3} & \textbf{Edge 4} & \textbf{Edge 5} & \textbf{Edge 6} \\
\textbf{Study} & \textbf{Kinematic variable} & \textbf{Bin edge 1} & \textbf{Bin edge 2} & \textbf{Bin edge 3} & \textbf{Bin edge 4} & \textbf{Bin edge 5} & \textbf{Bin edge 6} \\
\midrule
\multirow{4}{*}{$t\bar{t}t\bar{t}$}
& $p_T$                    & $0$     & $10.64$  & $11.10$ & $11.57$ & $12.17$ & $+\infty$ \\
& $\eta$                   & $0$     & $0.3444$ & $0.7052$ & $1.109$ & $1.633$ & $+\infty$ \\
& $\phi$/$\phi_{E_T^{\mathrm{miss}}}$ & $-\pi$  & $-3\pi/5$ & $-\pi/5$ & $\pi/5$ & $3\pi/5$ & $\pi$ \\
& $\|E_T^{\mathrm{miss}}\|$            & $0$     & $11.55$  & $12.03$ & $12.40$ & $12.76$ & $+\infty$ \\
\addlinespace
\midrule
\addlinespace
\multirow{4}{*}{SUSY $g\tilde{g}$}
& $p_T$                    & $0$     & $10.37$  & $10.76$ & $11.13$ & $11.52$ & $+\infty$ \\
& $\eta$                   & $0$     & $0.3877$ & $0.7918$ & $1.239$ & $1.789$ & $+\infty$ \\
& $\phi$/$\phi_{E_T^{\mathrm{miss}}}$ & $-\pi$  & $-3\pi/5$ & $-\pi/5$ & $\pi/5$ & $3\pi/5$ & $\pi$ \\
& $\|E_T^{\mathrm{miss}}\|$            & $0$     & $11.08$  & $11.31$ & $11.52$ & $11.78$ & $+\infty$ \\
\bottomrule
\end{tabular}
}
\end{table}

%\vspace{2cm}
% \begin{table}[h!]
% \caption{Bin definition for the Look-up table K=6.}
% \label{tab:binK6}
% \centering
% \scalebox{0.89}{
% \centering
% \begin{tabular}{c c c c c c c c c c}
% \toprule
% \textbf{study} & \textbf{kinematic} & \textbf{bin edge}  & \textbf{bin edge} & \textbf{bin edge} & \textbf{bin edge} &\textbf{bin edge} &\textbf{bin edge} &\textbf{bin edge}\\ %\cline{1-9}
% \specialrule{0.75pt}{1pt}{1pt}
% \multirow{4}{*}{\textbf{$t\bar{t}t\bar{t}$}} & $p_T$ & $0$ & $10.56$ & $10.94$ & $11.33$ & $11.75$ & $12.31$ &$+\infty$\\ \cline{2-9}
%  & $\eta$ & $0$ & $0.2853$ & $0.5806$ & $0.9$ & $1.26$  & $1.74$ & $+\infty$\\ \cline{2-9}
%  & $\phi$/$\phi_{E^{miss}_T}$ & $-\pi$ & $-2\pi/3$ & $-\pi/3$ & $0$ &$\pi/3$ & $2\pi/3$ & $\pi$ \\ \cline{2-9}
%  & $\|E^{miss}_T\|$ & $0$ & $11.43$ & $11.90$ & $12.22$ & $12.51$  & $12.83$ & $+\infty$\\ %\cline{1-9} 
% \specialrule{0.75pt}{1pt}{1pt}
% \multirow{4}{*}{\textbf{SUSY $g\tilde{g}$}} & $p_T$ & $0$ & $10.3$ & $10.64$ & $10.95$ & $11.25$ & $11.59$ & $+\infty$\\ \cline{2-9}
%  & $\eta$ & $0$ & $0.3228$ & $0.6551$ & $1.009$ & $1.403$ & $1.901$ & $+\infty$\\ \cline{2-9}
% & $\phi$/$\phi_{E^{miss}_T}$ & $-\pi$ & $-2\pi/3$ & $-\pi/3$ & $0$ &$\pi/3$ & $2\pi/3$ & $\pi$ \\ \cline{2-9}
%  & $\|E^{miss}_T\|$ & $0$ & $11.04$ & $11.23$ & $11.42$ & $11.6$ & $11.84$ & $+\infty$\\ 
% \bottomrule
% \end{tabular}
% }
% \end{table}
\vspace{4cm}
\begin{table}[htb!]
\centering
\caption{Bin definitions for the look-up table with $N=6$.}
\label{tab:binK6}
\small
\setlength{\tabcolsep}{4pt}
\renewcommand{\arraystretch}{1.15}
\scalebox{0.81}{
\begin{tabular}{lcccccccc}
\toprule
\textbf{Study} & \textbf{Kinematic variable} & \textbf{Bin edge 1} & \textbf{Bin edge 2} & \textbf{Bin edge 3} & \textbf{Bin edge 4} & \textbf{Bin edge 5} & \textbf{Bin edge 6} & \textbf{Bin edge 7} \\
\midrule
\multirow{4}{*}{$t\bar{t}t\bar{t}$}
& $p_T$                    & $0$     & $10.56$  & $10.94$ & $11.33$ & $11.75$ & $12.31$ & $+\infty$ \\
& $\eta$                   & $0$     & $0.2853$ & $0.5806$ & $0.9$   & $1.26$  & $1.74$  & $+\infty$ \\
& $\phi$/$\phi_{E_T^{\mathrm{miss}}}$ & $-\pi$  & $-2\pi/3$ & $-\pi/3$ & $0$    & $\pi/3$ & $2\pi/3$ & $\pi$ \\
& $\|E_T^{\mathrm{miss}}\|$            & $0$     & $11.43$  & $11.90$ & $12.22$ & $12.51$ & $12.83$ & $+\infty$ \\
\addlinespace
\midrule
\addlinespace
\multirow{4}{*}{SUSY $g\tilde{g}$}
& $p_T$                    & $0$     & $10.3$   & $10.64$ & $10.95$ & $11.25$ & $11.59$ & $+\infty$ \\
& $\eta$                   & $0$     & $0.3228$ & $0.6551$ & $1.009$ & $1.403$ & $1.901$ & $+\infty$ \\
& $\phi$/$\phi_{E_T^{\mathrm{miss}}}$ & $-\pi$  & $-2\pi/3$ & $-\pi/3$ & $0$    & $\pi/3$ & $2\pi/3$ & $\pi$ \\
& $\|E_T^{\mathrm{miss}}\|$            & $0$     & $11.04$  & $11.23$ & $11.42$ & $11.6$  & $11.84$ & $+\infty$ \\
\bottomrule
\end{tabular}

}
\end{table}

% \newpage
% \clearpage

\vspace{5cm}
%%%%%%%%%%%%%%%%%%%%%%%%%%%%%%%%%%
\section{Visualisation of the tokenization}
\label{app:Visualisation}
%%%%%%%%%%%%%%%%%%%%%%%%%%%%%%%%%%%
After tokenization, each event is represented as a sequence of tokens. Figures~\ref{fig:EncodingBin} and \ref{fig:EncodingVQ} (Figures~\ref{fig:EncodingBin-BSM} and \ref{fig:EncodingVQ-BSM}) show, for the 4-top (BSM) benchmark, the normalized token occurrence distributions for signal and background, along with their difference, for the look-up table and VQ-VAE tokenization schemes, respectively.

%\PM{Text referring to the plots, together with a short description, should be added here.}

\setcounter{figure}{0}   
\renewcommand{\thefigure}{C.\arabic{figure}}
\begin{figure}[h!]
    \centering
    \begin{subfigure}{\textwidth}
        \centering
        \includegraphics[width=0.99\linewidth]{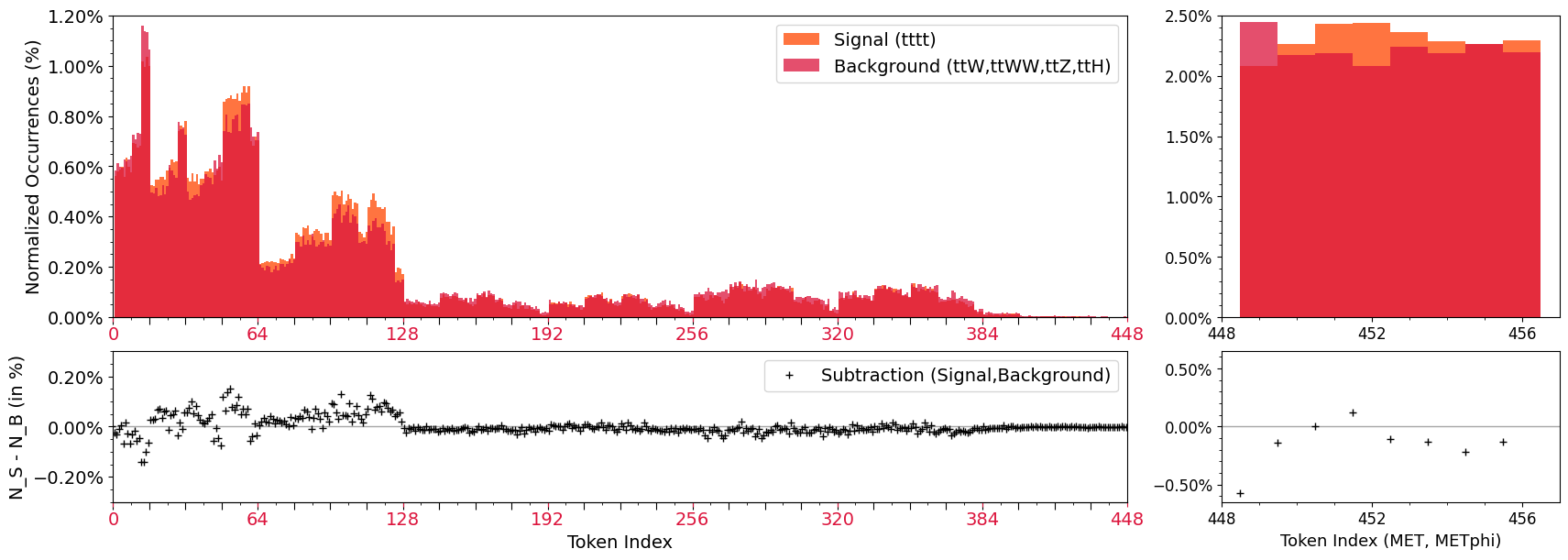}
        \caption{For a vocabulary size of 456.}
        \label{fig:enc456}
    \end{subfigure}
    \begin{subfigure}{\textwidth}
        \centering
        \includegraphics[width=0.99\linewidth]{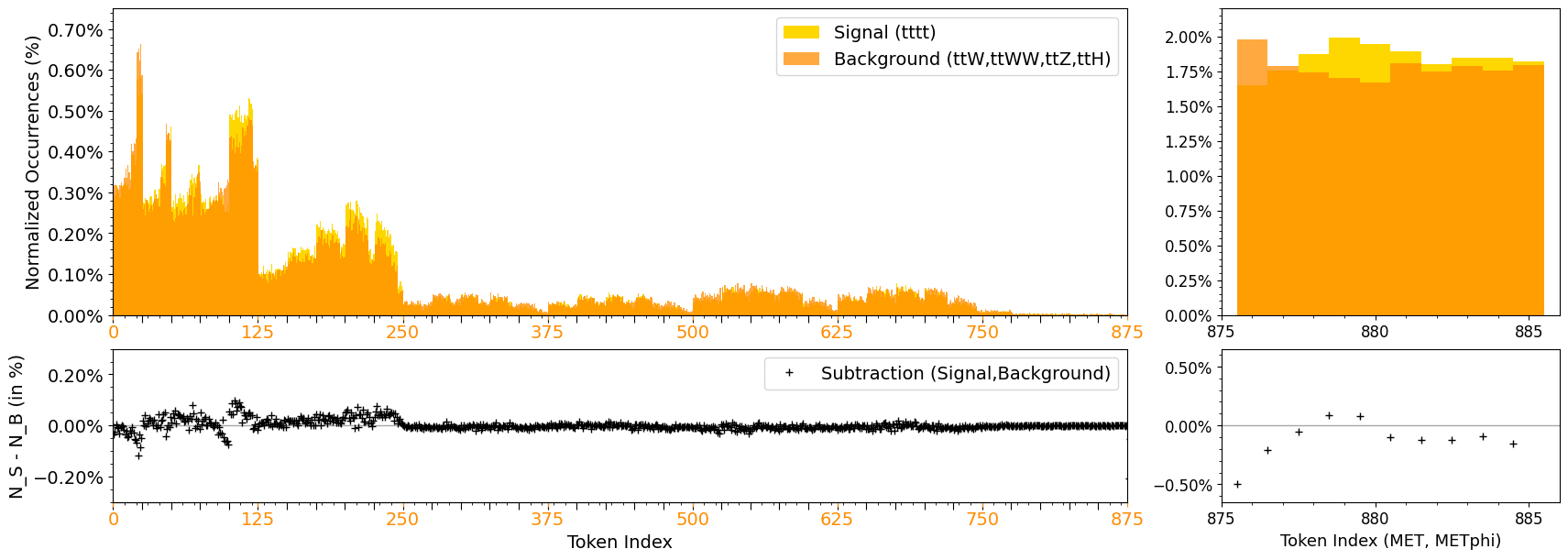}
        \caption{For a vocabulary size of 885.}
        \label{fig:enc885}
    \end{subfigure}
    \begin{subfigure}{\textwidth}
        \centering
        \includegraphics[width=0.99\linewidth]{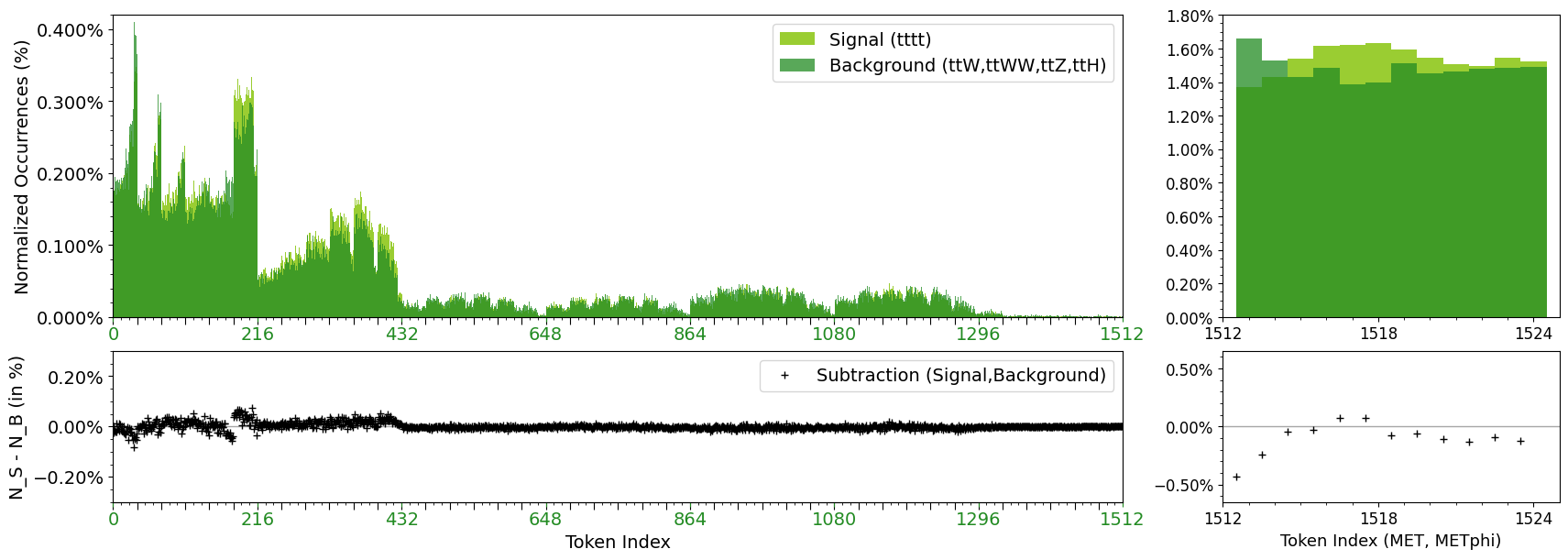}
        \caption{For a vocabulary size of 1524.}
        \label{fig:enc1524}
    \end{subfigure}
    % \caption{For the LUT tokenization scheme, distribution of token indices for \textbf{signal $t\bar{t}t\bar{t}$} (salmon in Fig.~\ref{fig:enc456}; yellow in Fig.~\ref{fig:enc885}; light green in Fig.~\ref{fig:enc1524}) and \textbf{background} (red in Fig.~\ref{fig:enc456}; orange in Fig.~\ref{fig:enc885}; dark green in Fig.~\ref{fig:enc1524}) events. Each histogram is normalized to the total number of non‑zero tokens within the corresponding class (signal or background). Lower panels display the subtraction (signal minus background) of the normalized occurrences per token index.} 
    \caption{For the LUT tokenization scheme, distributions of token indices for \textbf{signal} $t\bar{t}t\bar{t}$ events and \textbf{background} events, shown for vocabulary sizes of 456 (top), 885 (middle), and 1524 (bottom) %\PM{Why does the plot legend show 1524 instead of 1525?}
    . The signal distributions are shown in salmon in Fig.~\ref{fig:enc456}, yellow in Fig.~\ref{fig:enc885}, and light green in Fig.~\ref{fig:enc1524}, while the corresponding background distributions are shown in red, orange, and dark green. Each histogram is normalized to the total number of non-zero tokens within the corresponding class. The lower panels show the difference between the normalized signal and background occurrences as a function of token index. The right-hand insets highlight the token-index region corresponding to $E_T^{\mathrm{miss}}$ and $\phi_{E_T^{\mathrm{miss}}}$.}
    \label{fig:EncodingBin}
\end{figure}

\begin{figure}[p]
    \centering
    \begin{subfigure}{\textwidth}
        \centering
        \includegraphics[width=0.99\linewidth]{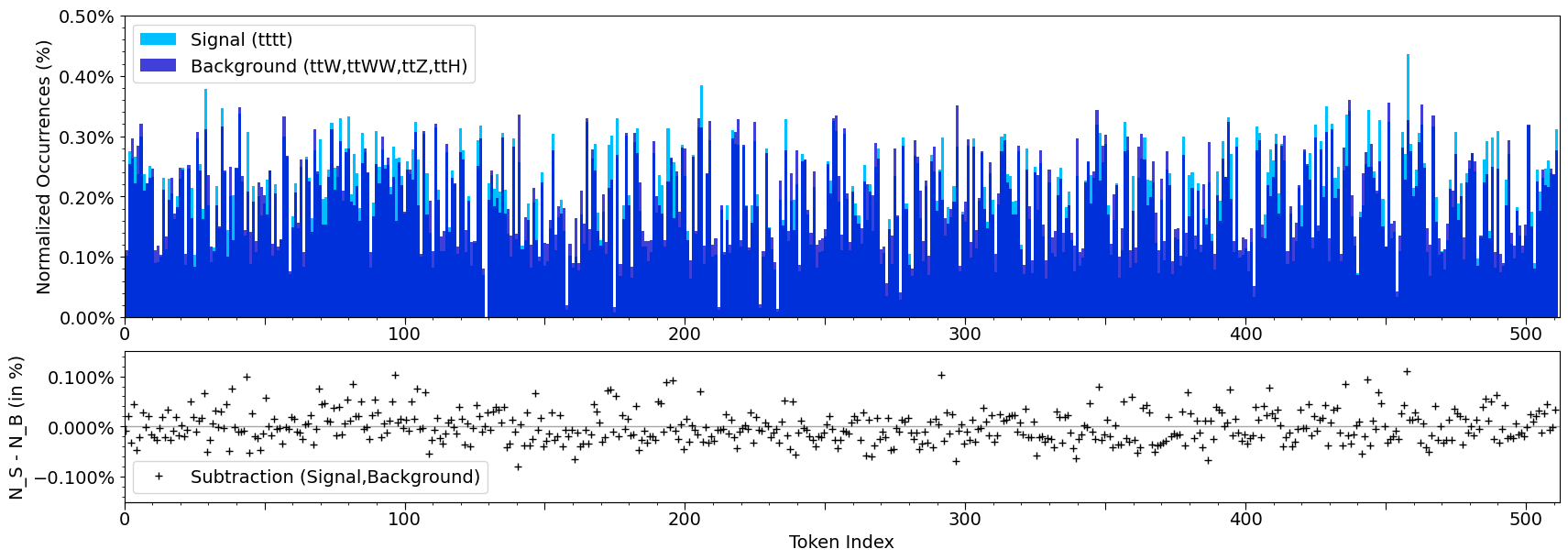}
        \caption{For a vocabulary size of 512.}
        \label{fig:enc512}
    \end{subfigure}
    \begin{subfigure}{\textwidth}
        \centering
        \includegraphics[width=0.99\linewidth]{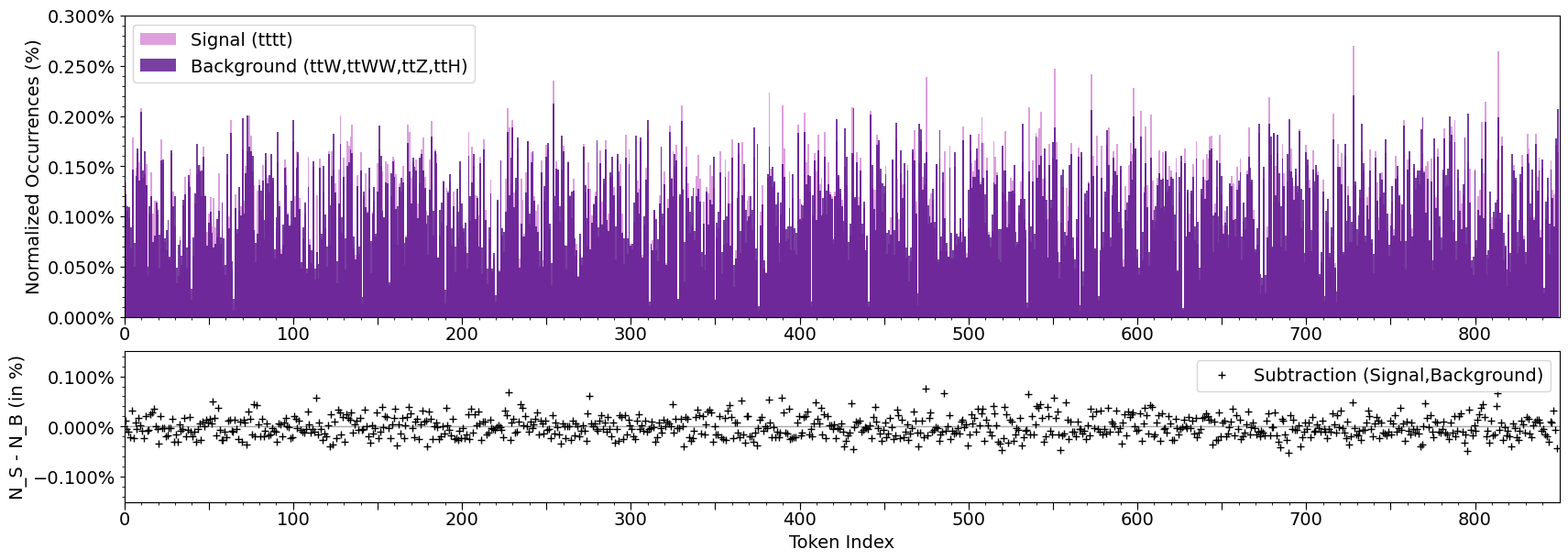}
        \caption{For a vocabulary size of 850.}
        \label{fig:enc850}
    \end{subfigure}
    \begin{subfigure}{\textwidth}
        \centering
        \includegraphics[width=0.99\linewidth]{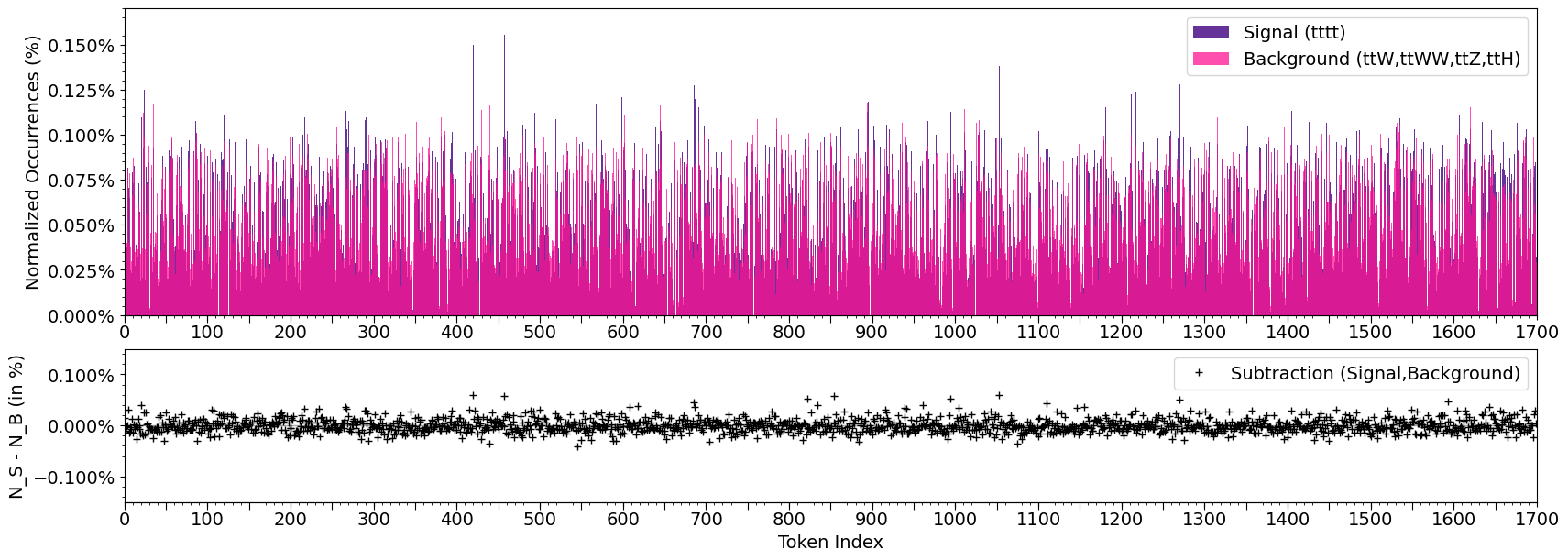}
        \caption{For a vocabulary size of 1700.}
        \label{fig:enc1700}
    \end{subfigure}
    \caption{For the VQ-VAE tokenization scheme, distributions of token indices for \textbf{signal} $t\bar{t}t\bar{t}$ events and \textbf{background} events, shown for vocabulary sizes of 512 (top), 850 (middle), and 1700 (bottom). The signal distributions are shown in cyan in Fig.~\ref{fig:enc512}, light violet in Fig.~\ref{fig:enc850}, and dark violet in Fig.~\ref{fig:enc1700}, while the corresponding background distributions are shown in blue, purple, and pink. Each histogram is normalized to the total number of non-zero tokens within the corresponding class. The lower panels show the difference between the normalized signal and background occurrences as a function of token index.}
    \label{fig:EncodingVQ}
\end{figure}

\begin{figure}[p]
    \centering
    \begin{subfigure}{\textwidth}
        \centering
        \includegraphics[width=0.99\linewidth]{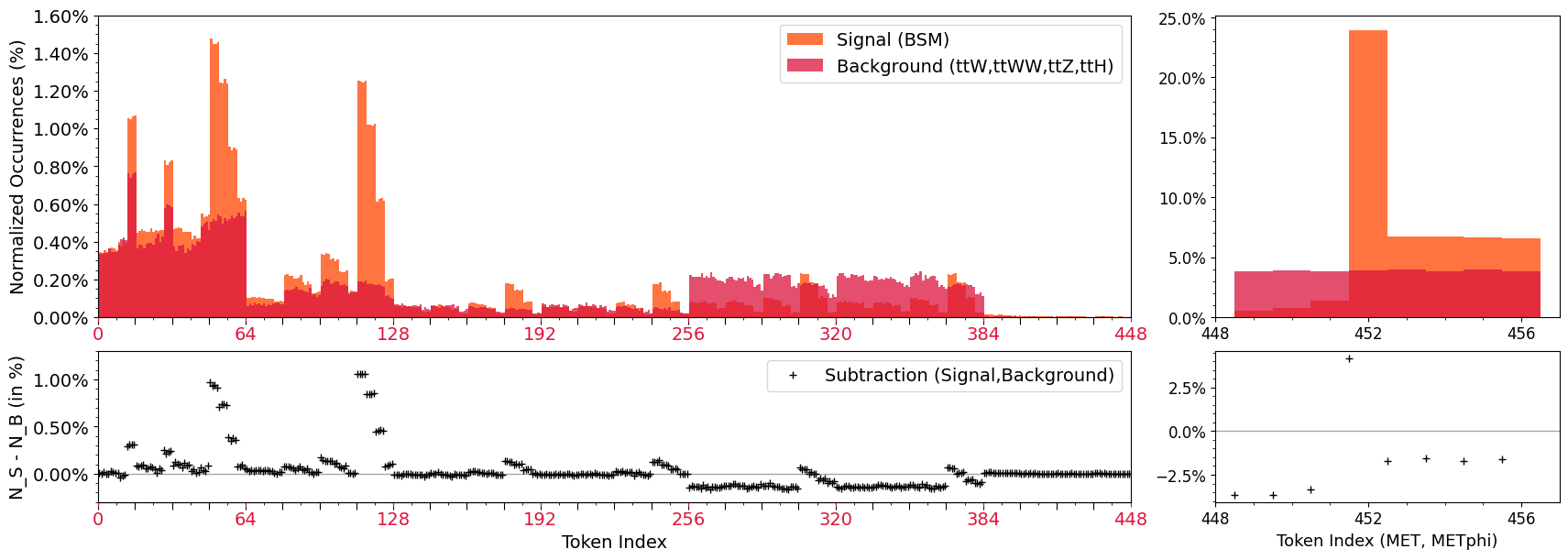}
        \caption{For a vocabulary size of 456.}
        \label{fig:enc456-BSM}
    \end{subfigure}
    \begin{subfigure}{\textwidth}
        \centering
        \includegraphics[width=0.99\linewidth]{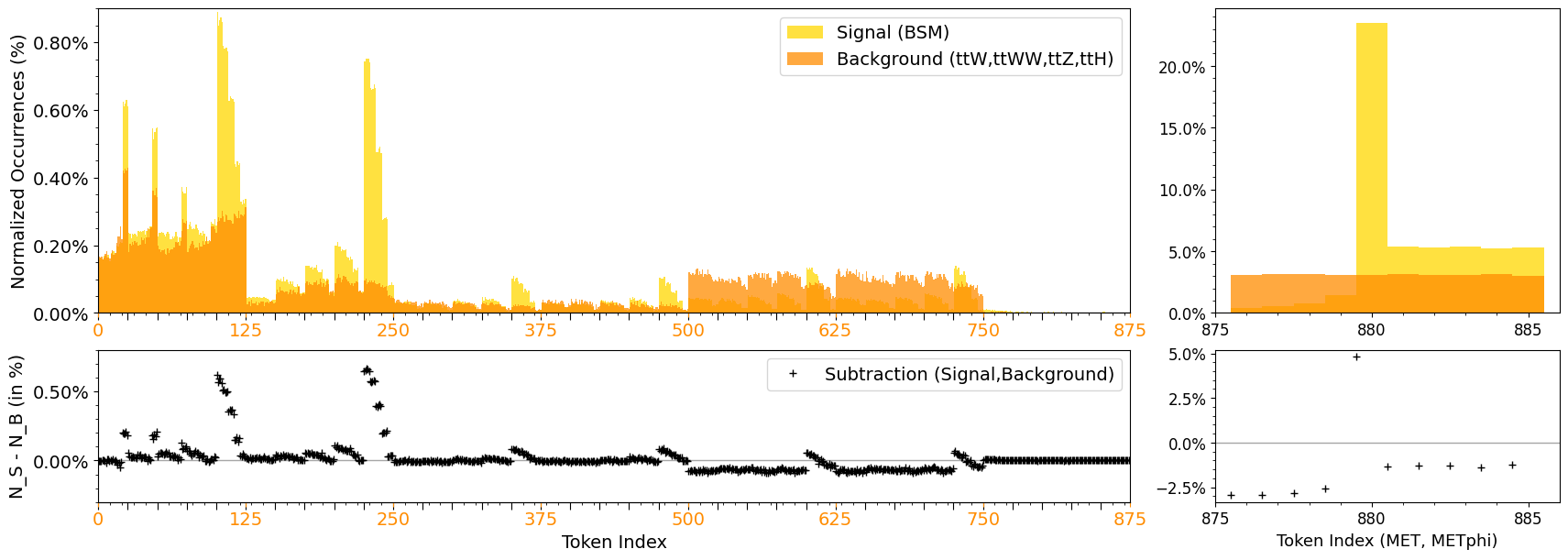}
        \caption{For a vocabulary size of 885.}
        \label{fig:enc885-BSM}
    \end{subfigure}
    \begin{subfigure}{\textwidth}
        \centering
        \includegraphics[width=0.99\linewidth]{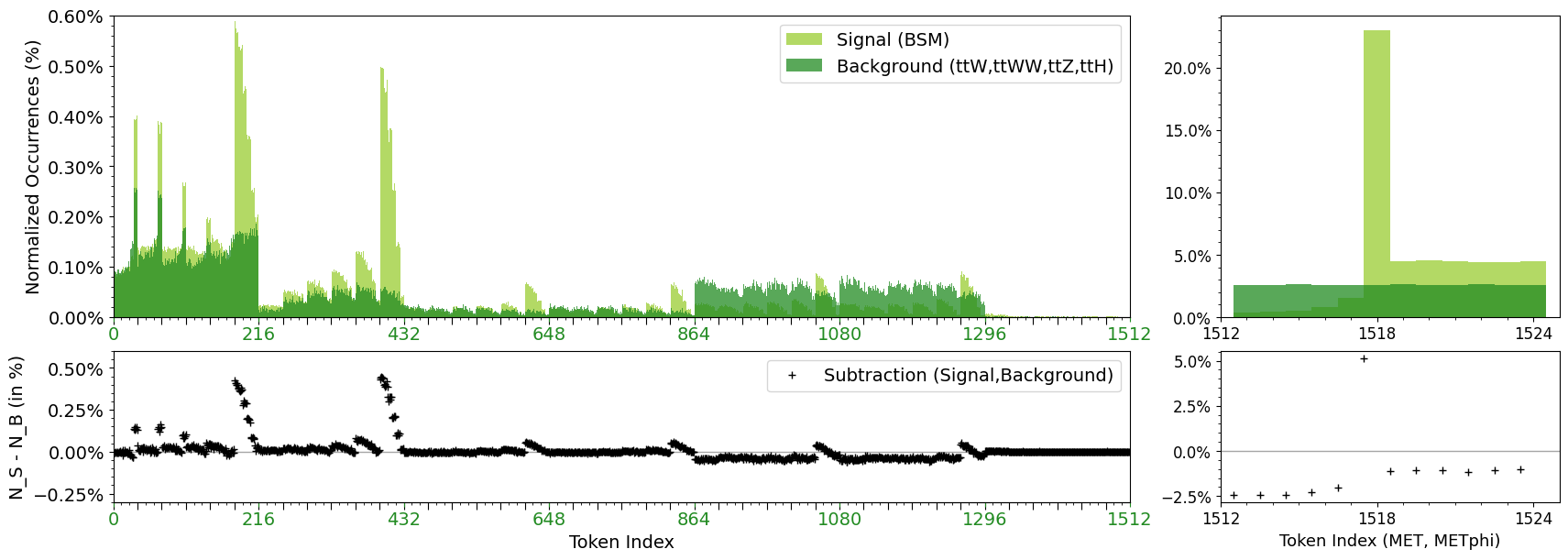}
        \caption{For a vocabulary size of 1524.}
        \label{fig:enc1524-BSM}
    \end{subfigure}
    \caption{For the LUT tokenization scheme, distributions of token indices for \textbf{BSM signal} and \textbf{background} events, shown for vocabulary sizes of 456 (top), 885 (middle), and 1524 (bottom). The signal distributions are shown in salmon in Fig.~\ref{fig:enc456-BSM}, yellow in Fig.~\ref{fig:enc885-BSM}, and light green in Fig.~\ref{fig:enc1524-BSM}, while the corresponding background distributions are shown in red, orange, and dark green. Each histogram is normalized to the total number of non-zero tokens within the corresponding class. The lower panels show the difference between the normalized signal and background occurrences as a function of token index. The right-hand insets highlight the token-index region corresponding to $E_T^{\mathrm{miss}}$ and $\phi_{E_T^{\mathrm{miss}}}$ for the SUSY benchmark.}
    \label{fig:EncodingBin-BSM}
\end{figure}

\begin{figure}[p]
    \centering
    \begin{subfigure}{\textwidth}
        \centering
        \includegraphics[width=0.99\linewidth]{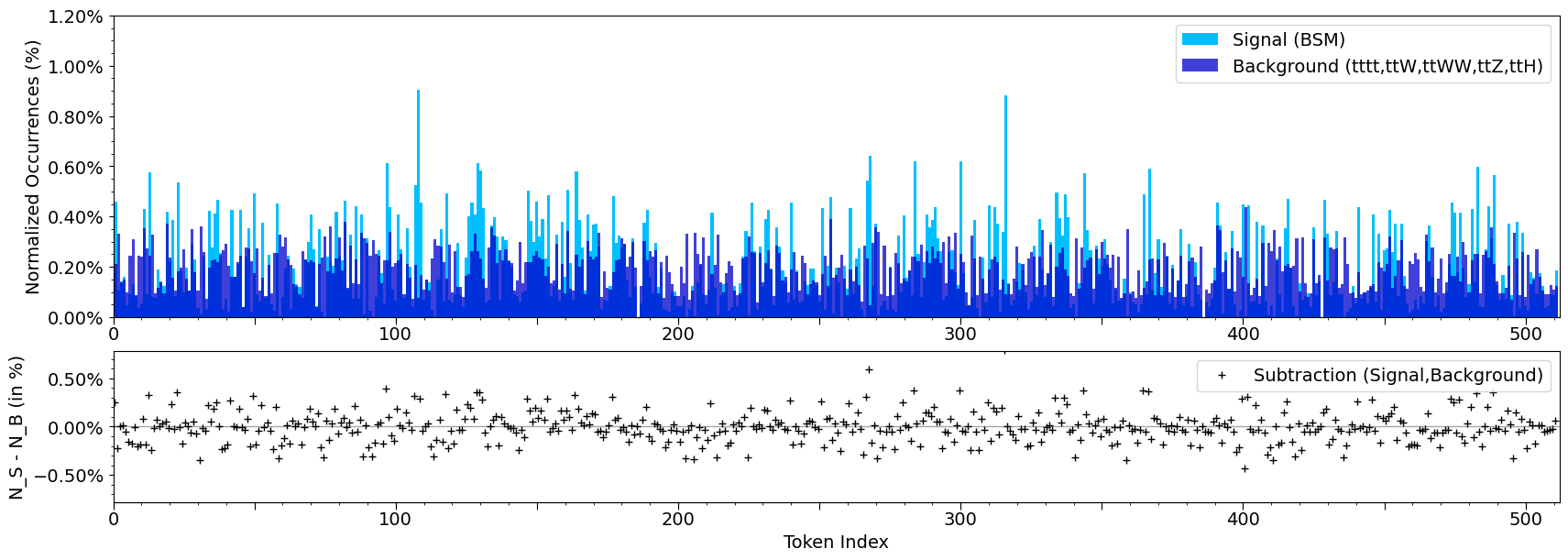}
        \caption{For a vocabulary size of 512.}
        \label{fig:enc512-BSM}
    \end{subfigure}
    \begin{subfigure}{\textwidth}
        \centering
        \includegraphics[width=0.99\linewidth]{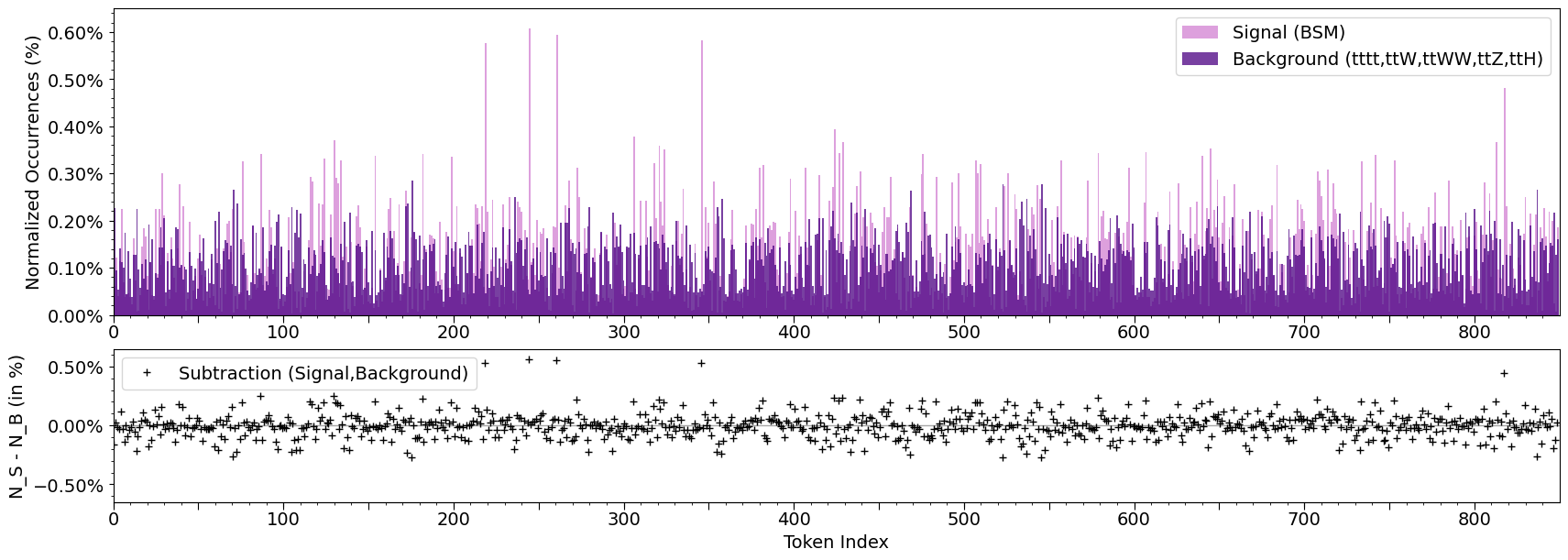}
        \caption{For a vocabulary size of 850.}
        \label{fig:enc850-BSM}
    \end{subfigure}
    \begin{subfigure}{\textwidth}
        \centering
        \includegraphics[width=0.99\linewidth]{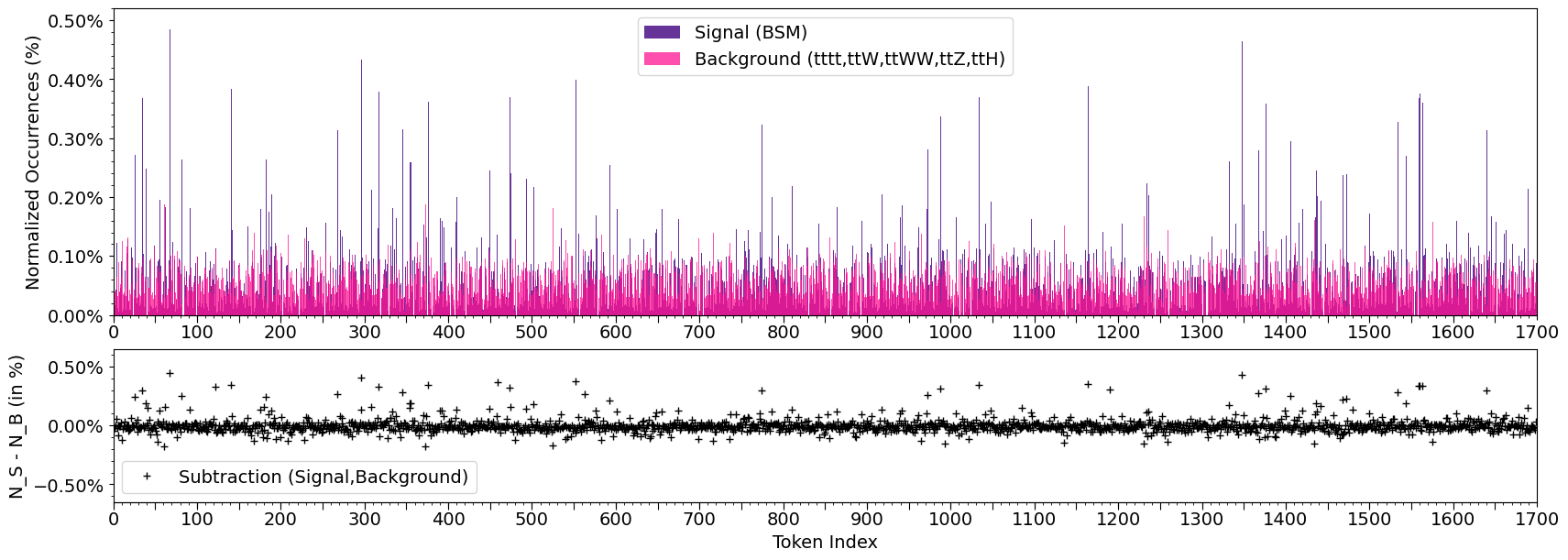}
        \caption{For a vocabulary size of 1700.}
        \label{fig:enc1700-BSM}
    \end{subfigure}
    \caption{For the VQ-VAE tokenization scheme, distributions of token indices for \textbf{BSM signal} events and \textbf{background} events, shown for vocabulary sizes of 512 (top), 850 (middle), and 1700 (bottom). The signal distributions are shown in cyan in Fig.~\ref{fig:enc512-BSM}, light violet in Fig.~\ref{fig:enc850-BSM}, and dark violet in Fig.~\ref{fig:enc1700-BSM}, while the corresponding background distributions are shown in blue, purple, and pink. Each histogram is normalized to the total number of non-zero tokens within the corresponding class. The lower panels show the difference between the normalized signal and background occurrences as a function of token index.}
    \label{fig:EncodingVQ-BSM}
\end{figure}

\clearpage

%%%%%%%%%%%%%%%%%%%%%%%%%%%%%%%%%%
\section{Hyperparameters of downstream models}
\label{app:HyperparametersLLM}
%%%%%%%%%%%%%%%%%%%%%%%%%%%%%%%%%%

%\PM{Text referring to the tables, together with a short description, should be added here.}
For each tokenization strategy, we scan several hyperparameters for the downstream masked-token prediction model. The values corresponding to the  best-performing models obtained for the $t\bar{t}t\bar{t}$ and the SUSY $g\tilde{g}$ benchmarks are listed in Tables~\ref{tab:downstream_hyperparameters_4top} and \ref{tab:downstream_hyperparameters_BSM}, respectively. 

\vspace{2cm}
\setcounter{table}{0}   
\renewcommand{\thetable}{D.\arabic{table}}
% \begin{table}[htbp]
\begin{table}[htb!]
\centering
\caption{Hyperparameters of the best-performing masked-token prediction models for each tokenization strategy in the four-top study.}%\PM{Should this be 850 instead of 875?}}
\label{tab:downstream_hyperparameters_4top}
\small
\setlength{\tabcolsep}{5pt}
\renewcommand{\arraystretch}{1.15}
\scalebox{0.9}{
\begin{tabular}{lcccccc}
\toprule
& \multicolumn{3}{c}{\textbf{LUT}} & \multicolumn{3}{c}{\textbf{VQ-VAE}} \\
\cmidrule(lr){2-4} \cmidrule(lr){5-7}
\textbf{Vocabulary size} & \textbf{456} & \textbf{885} & \textbf{1524} & \textbf{512} & \textbf{850} & \textbf{1700} \\
\midrule
\multicolumn{7}{l}{\textit{Architecture}} \\
Number of transformer layers & 2 & 2 & 2 & 2 & 2 & 2 \\
Number of attention heads    & 4 & 4 & 4 & 4 & 4 & 4 \\
Input size                   & 20 & 20 & 20 & 20 &  & 20 \\
Hidden size                  & 64 & 64 & 64 &  & 64 & 64 \\
FFN expansion factor         & 1 & 1 & 1 & 1 & 1 & 1  \\
Positional encoding layer    & none & none & rotary & none  & none & rotary \\
Dropout rate                 & 0.05 & 0.05 & 0.05 & 0.05 & 0.05 & 0.05 \\
\addlinespace
\multicolumn{7}{l}{\textit{Training}} \\
Batch size                   & 512 & 512  & 512 & 512 & 512 & 512 \\
Peak learning rate (Adam)    & $5.10^{-05}$ & $5.10^{-05}$ & none  & none & none &  none\\
Total LR steps (Adam)        & 10000 & 20000 & --  & -- & -- & -- \\
Warm-up LR steps (Adam)      & 1000 & 2000 & -- & -- & -- & -- \\
Gradient clipping norm       & 1000 & 1000 & 1000 & 1000 & 1000 & 1000 \\
Number of training epochs    & 40 & 49 & 20 & 33 & 35 & 67 \\
\addlinespace
\multicolumn{7}{l}{\textit{Early stopping}} \\
Minimum loss improvement     & 0.1\% & 0.1\% & 0.1\% & 0.1\% & 0.1\% & 0.1\% \\
Patience                     & 5 & 5 & 5 & 5 & 5 & 5 \\
Validation-training loss divergence & 0.2 & 0.2 & 0.2 & 0.2 & 0.2 & 0.2 \\
\bottomrule
\end{tabular}
}
\end{table}

% \begin{table}[htbp]
\begin{table}[htb!]
\centering
\caption{Hyperparameters of the best-performing masked-token prediction models for each tokenization strategy in the BSM study.}%\PM{or transformer-based rather than LLM-like?} models evaluated with each tokenization strategy in the BSM study. }
\label{tab:downstream_hyperparameters_BSM}
\small
\setlength{\tabcolsep}{5pt}
\renewcommand{\arraystretch}{1.15}
\scalebox{0.9}{
\begin{tabular}{lcccccc}
\toprule
& \multicolumn{3}{c}{\textbf{LUT}} & \multicolumn{3}{c}{\textbf{VQ-VAE}} \\
\cmidrule(lr){2-4} \cmidrule(lr){5-7}
\textbf{Vocabulary size} & \textbf{456} & \textbf{885} & \textbf{1524} & \textbf{512} & \textbf{850} & \textbf{1700} \\
\midrule
\multicolumn{7}{l}{\textit{Architecture}} \\
Number of transformer layers & 2 & 2 & 2 & 2 & 2 & 2 \\
Number of attention heads    & 4 & 4 & 4 & 4 & 4 & 4 \\
Input size                   & 20 & 20 & 20 & 20 & 20 & 20 \\
Hidden size                  & 64 & 64 & 64 & 64 & 64 & 64 \\
FFN expansion factor         & 2 & 2 & 1 & 1 & 1 & 1 \\
Positional encoding layer    & none & none  & none & sinusoidal & rotary & sinusoidal \\
Dropout rate                 & 0.05 & 0.05 & 0.05 & 0.05 & 0.05 & 0.1 \\
\addlinespace
\multicolumn{7}{l}{\textit{Training}} \\
Batch size                   & 512 & 512 &  512 & 512 & 512 & 512 \\
Peak learning rate (Adam)    & $5.10^{-05}$ & $5.10^{-05}$ & none & none & none & none \\
Total LR steps (Adam)        & 10000 & 10000 & -- & -- & -- & -- \\
Warm-up LR steps (Adam)      & 1000 & 1000 & --  & -- & -- & -- \\
Gradient clipping norm       & 5 & 5 & 5 & 1000 & 1000 & 5 \\
Number of training epochs    & 18 & 18 & 37 & 33 & 18 & 48 \\
\addlinespace
\multicolumn{7}{l}{\textit{Early stopping}} \\
Minimum loss improvement     & 0.01\% & 0.01\% & 0.01\% & 0.01\% & 0.01\% & 0.01\% \\
Patience                     & 5 & 5 & 5 & 5 & 5 & 5 \\
Validation-training loss divergence & 0.2 & 0.2 & 0.2 & 0.2 & 0.2 & 0.2 \\
\bottomrule
\end{tabular}
}
\end{table}

\clearpage

%%%%%%%%%%%%%%%%%%%%%%%%%%%%%%%%%%
\section{Anomaly score distributions}
\label{app:Histograms}
%%%%%%%%%%%%%%%%%%%%%%%%%%%%%%%%%%

At inference, the masked-token prediction model reconstructs each token of every \textit{test} event one at a time; performance is evaluated via sparse categorical cross-entropy (see Section~\ref{sec:LLManomalydetector}). These per-token losses are averaged to yield a single event-level reconstruction score, the anomaly score. Figures~\ref{fig:histo4top} and \ref{fig:histoBSM} show the resulting score distributions for signal and background events in the $t\bar{t}t\bar{t}$ and SUSY $g\tilde{g}$ benchmarks, respectively, across all tokenization schemes and vocabulary sizes. The optimal anomaly detection threshold and the overlap area between the two distributions are also indicated.

%\PM{Text referring to the plots, together with a short description, should be added here.}

\setcounter{figure}{0}   
\renewcommand{\thefigure}{E.\arabic{figure}}

\begin{figure}[htb!]
\centering

\begin{subfigure}{0.49\textwidth}
\includegraphics[width=\linewidth]{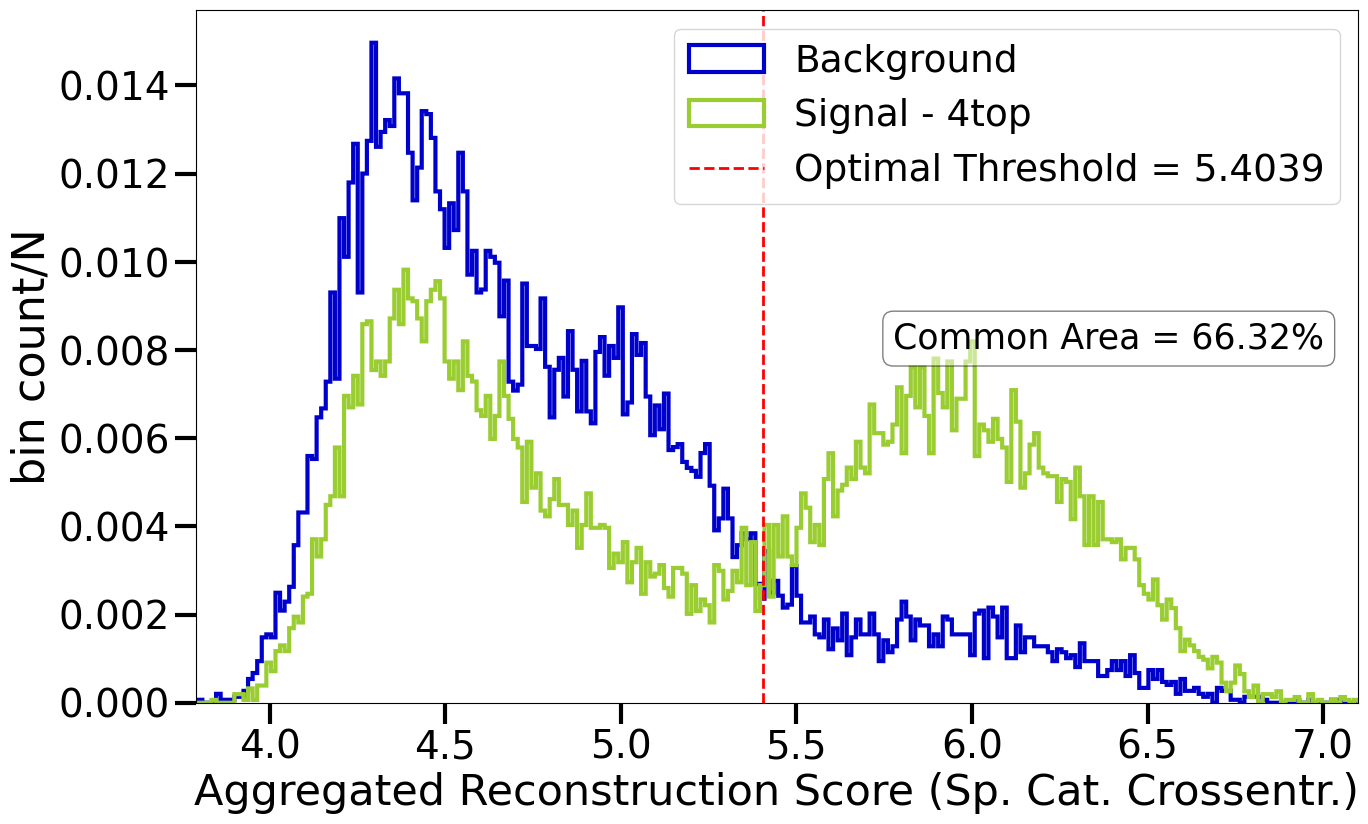} 
\caption{LUT tokenization, vocabulary size = 456.}
\label{fig:Histogram456}
\end{subfigure}
\hfill
\begin{subfigure}{0.49\textwidth}
\includegraphics[width=\linewidth]{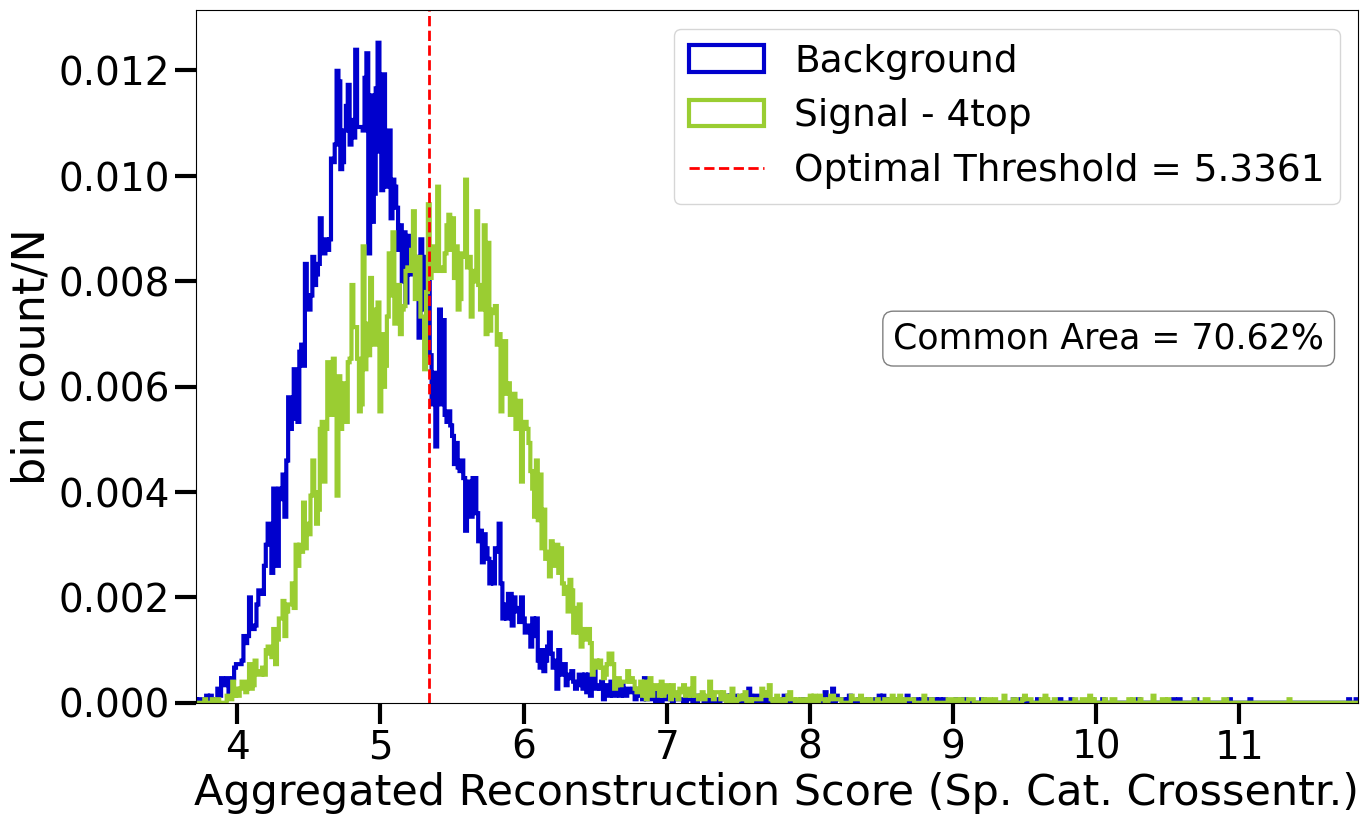} 
\caption{VQ-VAE tokenization, vocabulary size = 512.}
\label{fig:Histogram512}
\end{subfigure}

\vspace{0.5em}

\begin{subfigure}{0.49\textwidth}
\includegraphics[width=\linewidth]{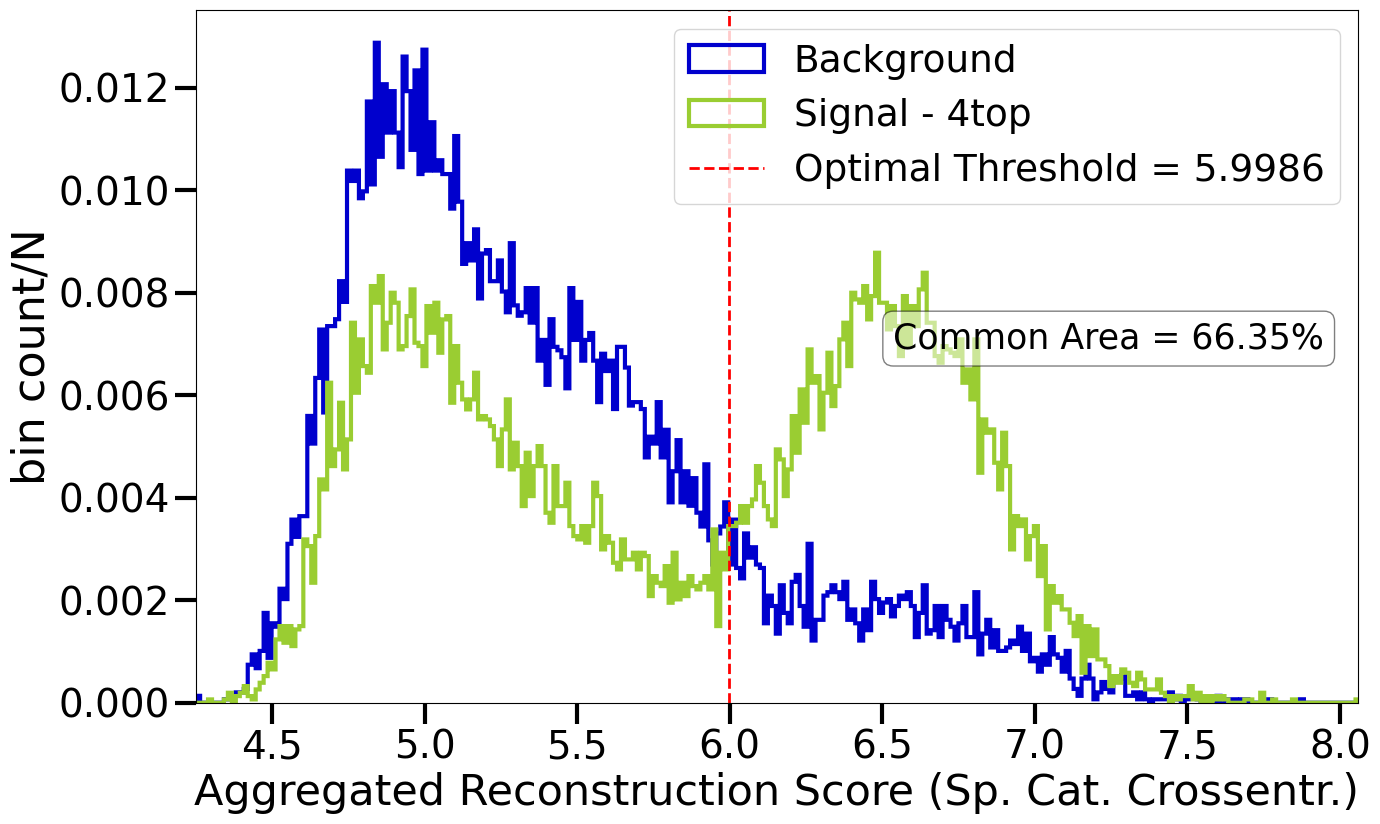} 
\caption{LUT tokenization, vocabulary size = 885.}
\label{fig:Histogram885}
\end{subfigure}
\hfill
\begin{subfigure}{0.49\textwidth}
\includegraphics[width=\linewidth]{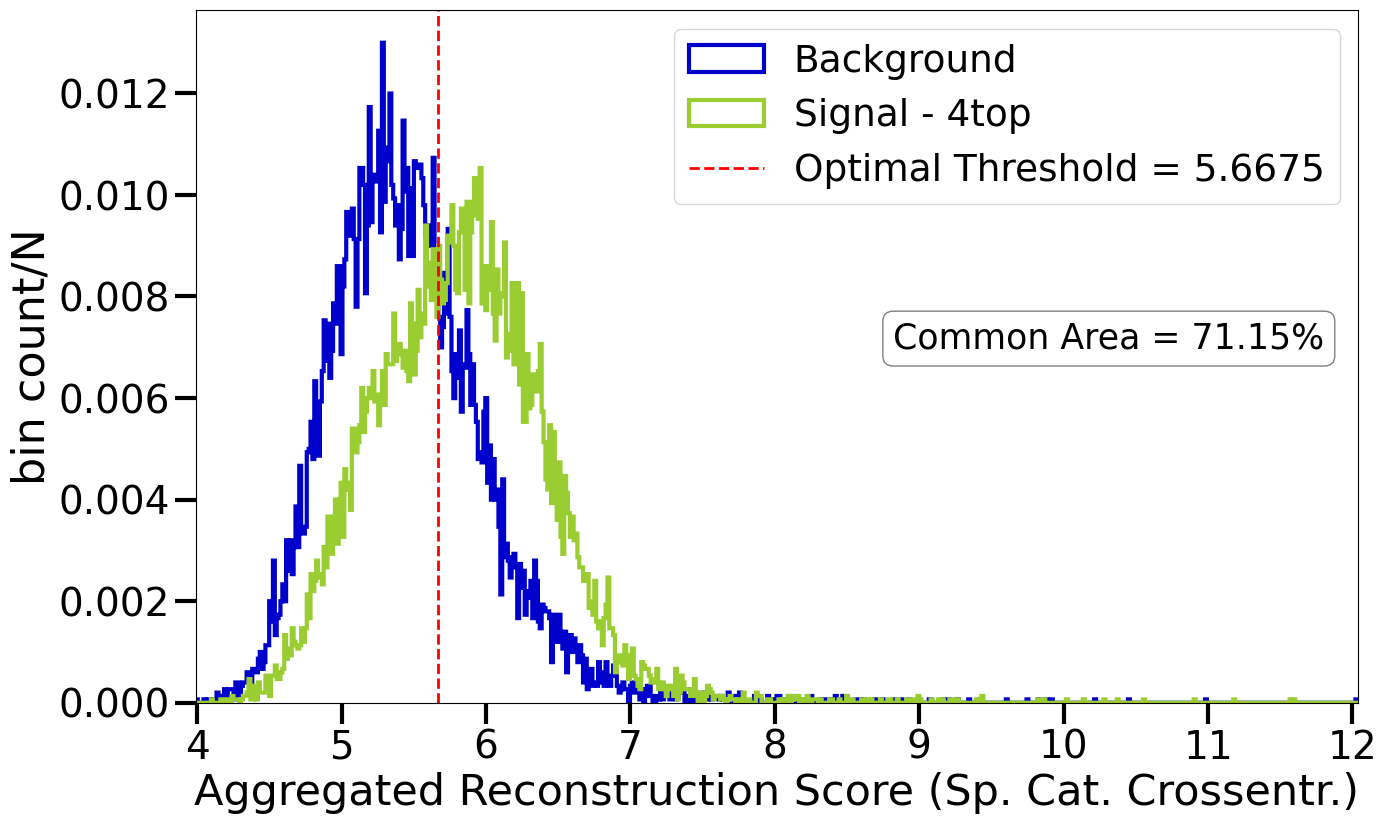} 
\caption{VQ-VAE tokenization, vocabulary size = 850.}
\label{fig:Histogram850}
\end{subfigure}

\vspace{0.5em}

\begin{subfigure}{0.49\textwidth}
\includegraphics[width=\linewidth]{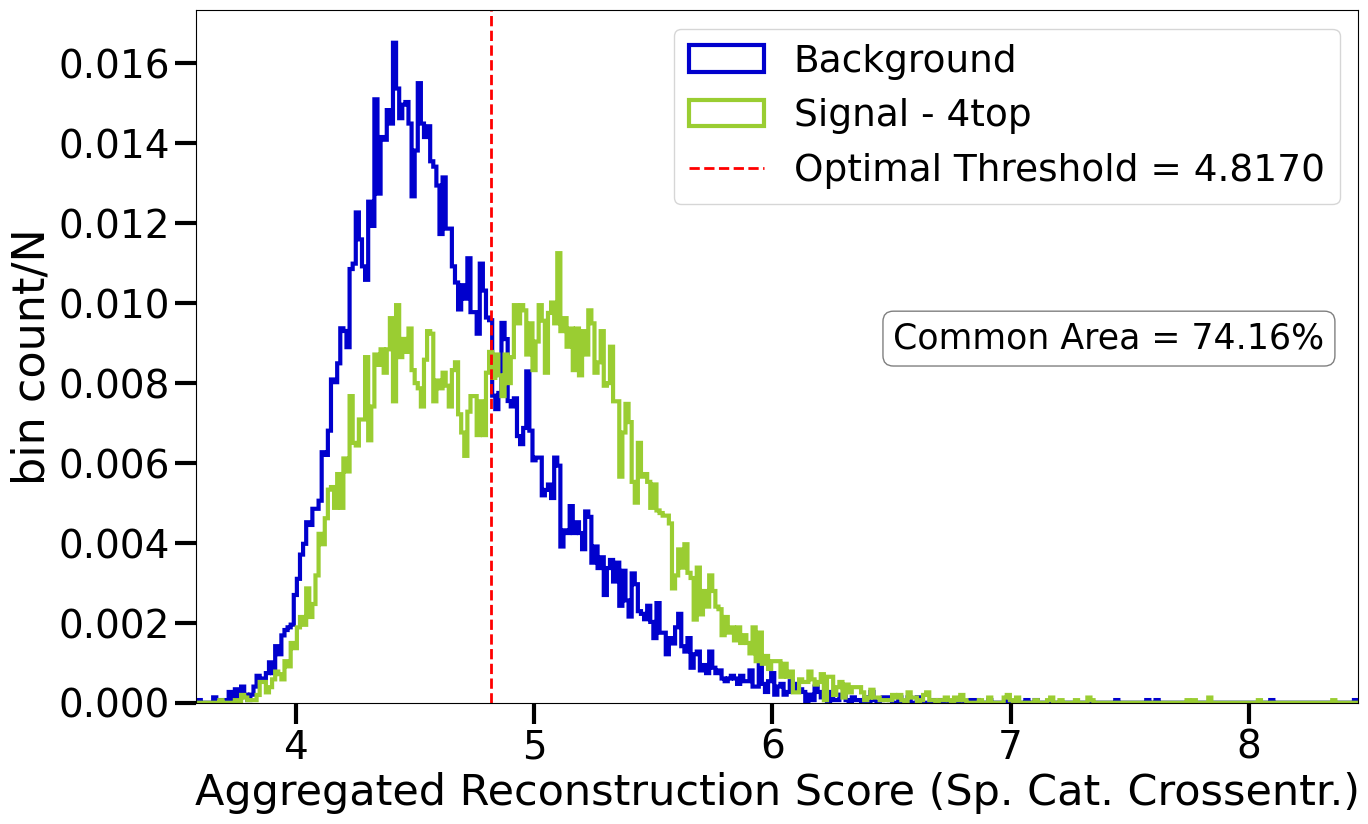} 
\caption{LUT tokenization, vocabulary size = 1524.}
\label{fig:Histogram1524}
\end{subfigure}
\hfill
\begin{subfigure}{0.49\textwidth}
\includegraphics[width=\linewidth]{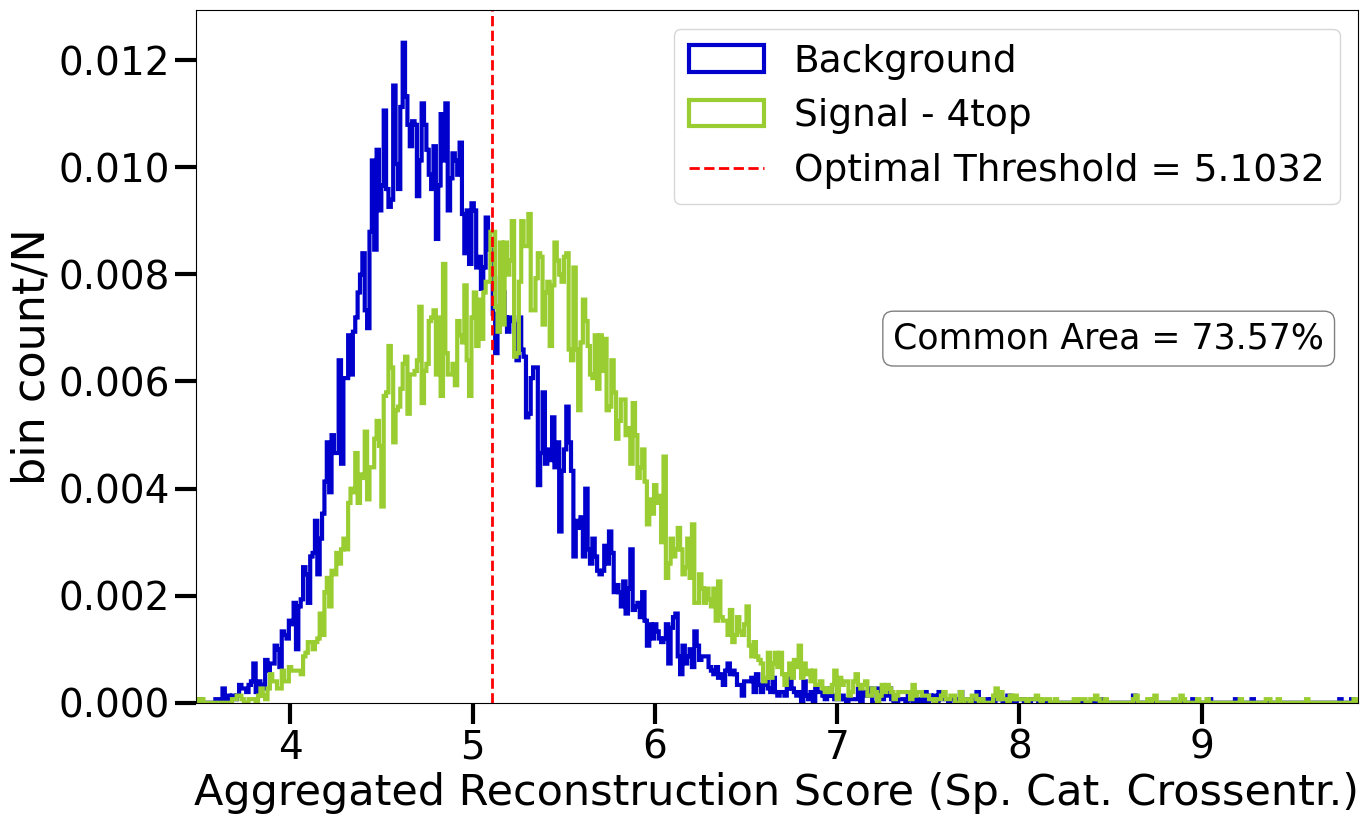} 
\caption{VQ-VAE  tokenization, vocabulary size = 1700.}
\label{fig:Histogram1700}
\end{subfigure}
\caption{Anomaly score distributions for background (blue) and signal (green) events in the $t\bar{t}t\bar{t}$ benchmark, as obtained from the downstream model. The optimal discrimination threshold is shown in red.}
\label{fig:histo4top}
\end{figure}

\begin{figure}[htb!]
\centering

\begin{subfigure}{0.49\textwidth}
\includegraphics[width=\linewidth]{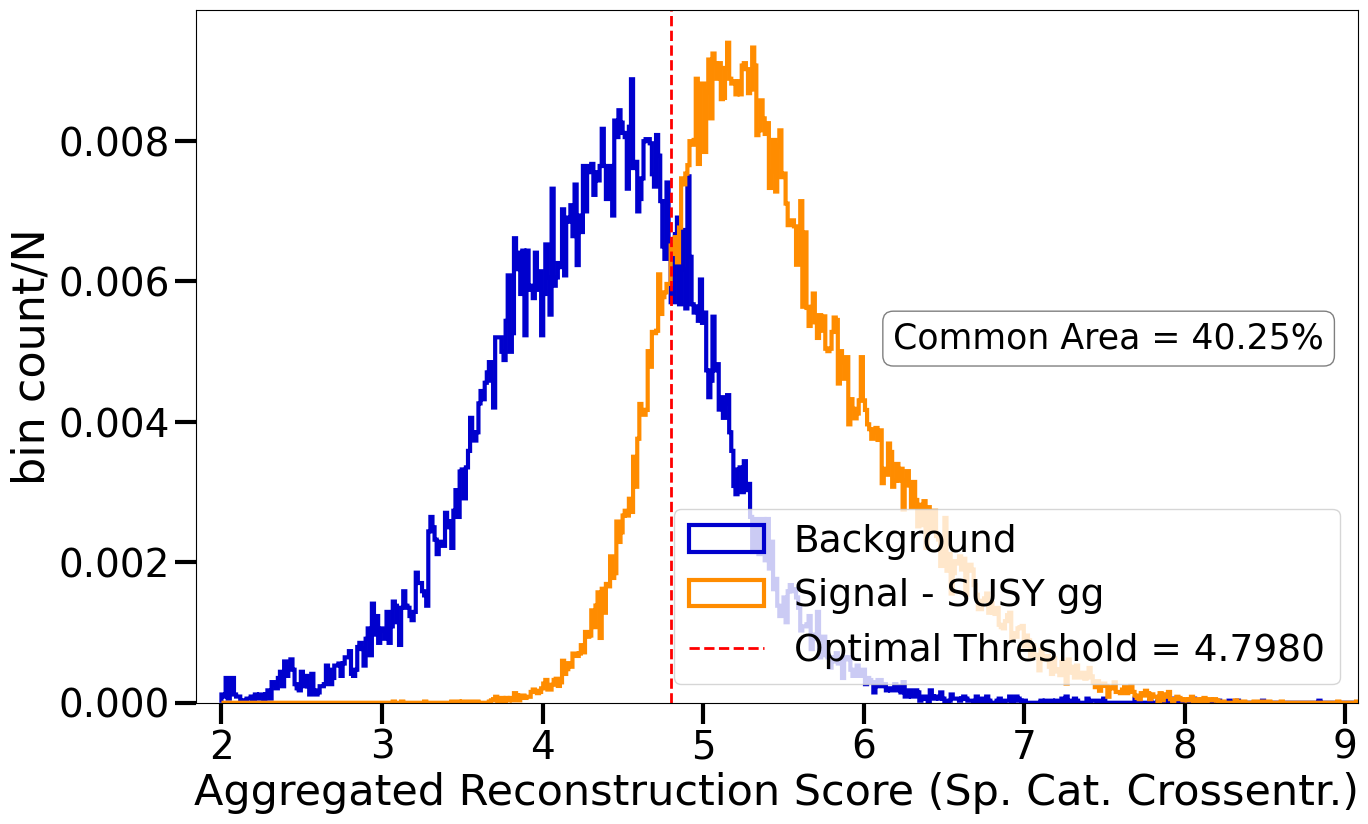} 
\caption{LUT tokenization, vocabulary size = 456.}
\label{fig:Histogram456-BSM}
\end{subfigure}
\hfill
\begin{subfigure}{0.49\textwidth}
\includegraphics[width=\linewidth]{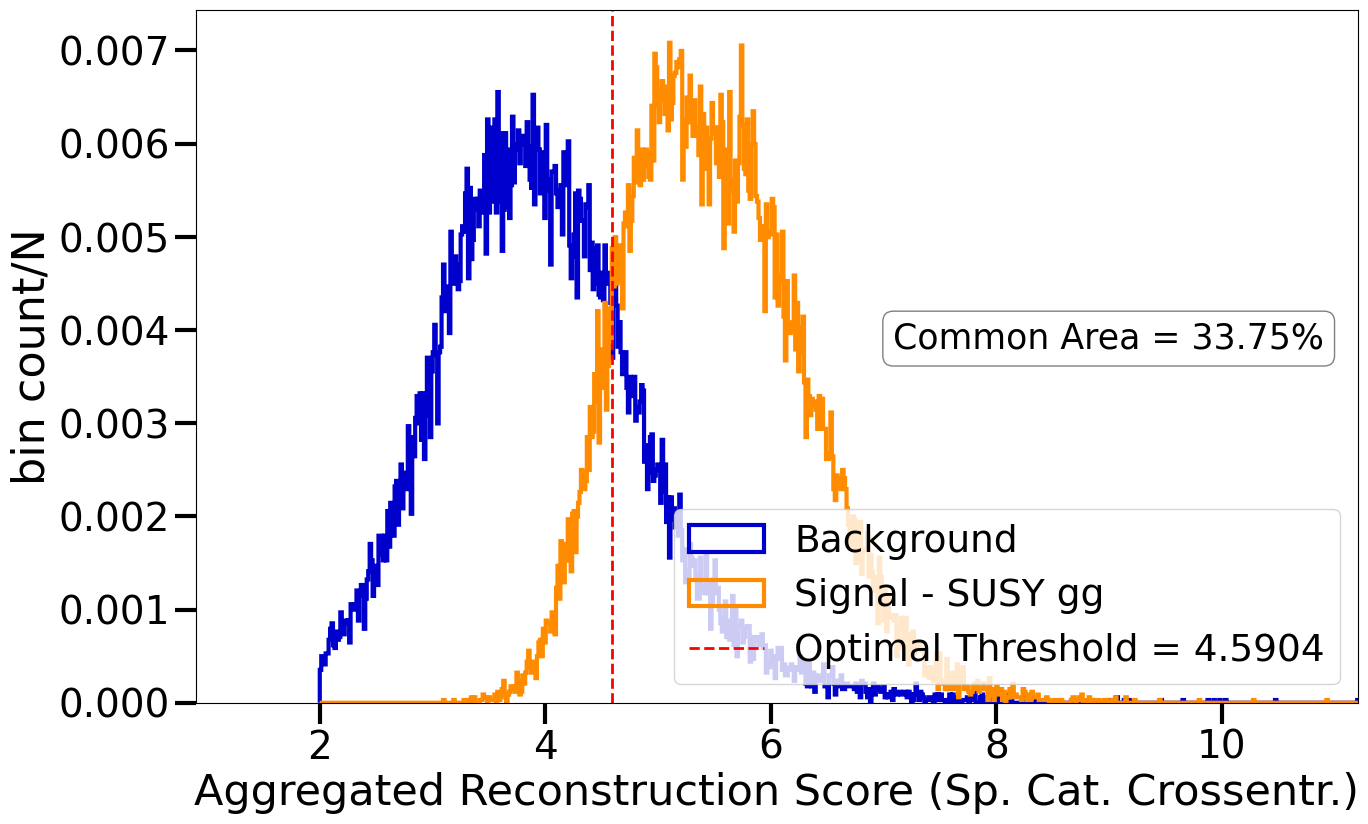} 
\caption{VQ-VAE tokenization, vocabulary size = 512.}
\label{fig:Histogram512-BSM}
\end{subfigure}

\vspace{0.5em}

\begin{subfigure}{0.49\textwidth}
\includegraphics[width=\linewidth]{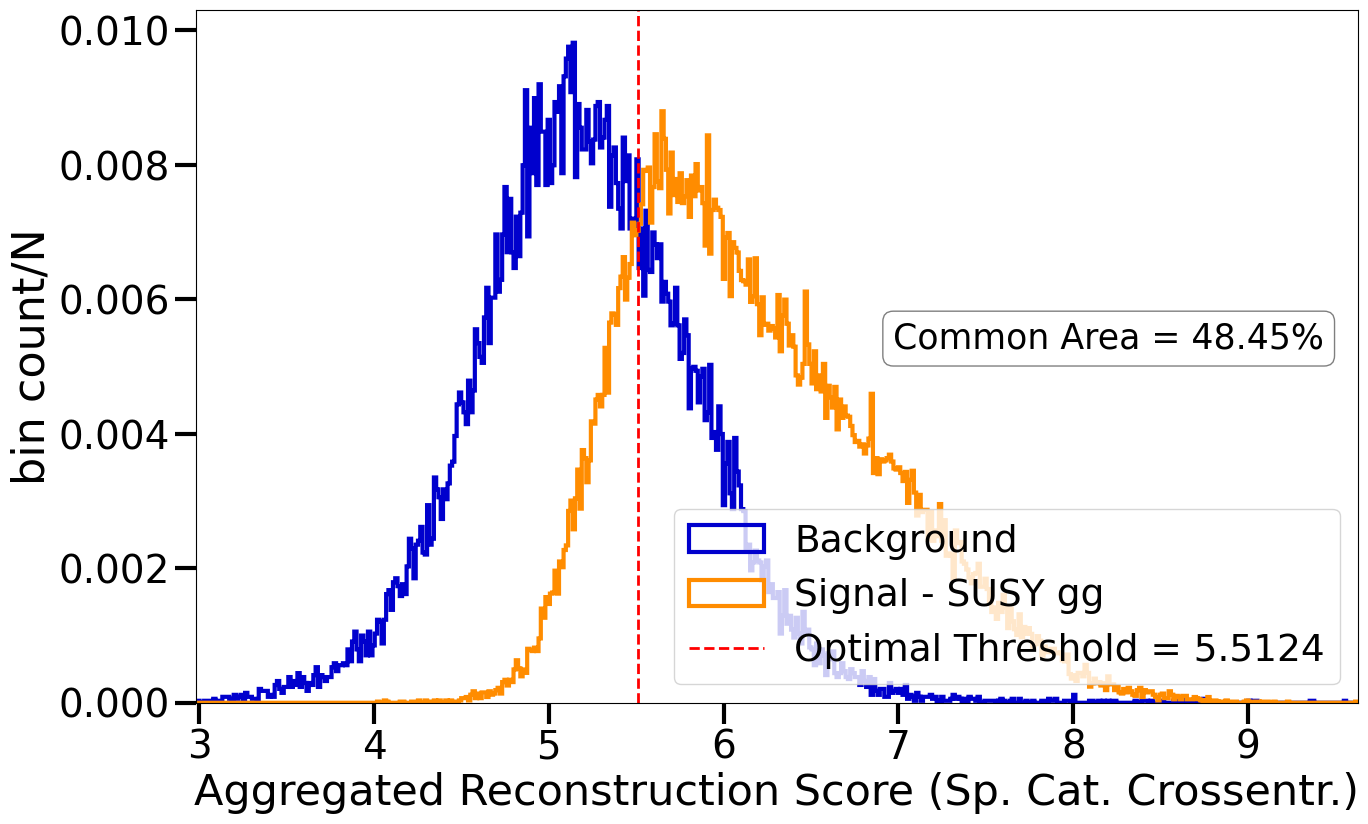} 
\caption{LUT tokenization, vocabulary size = 885.}
\label{fig:Histogram885-BSM}
\end{subfigure}
\hfill
\begin{subfigure}{0.49\textwidth}
\includegraphics[width=\linewidth]{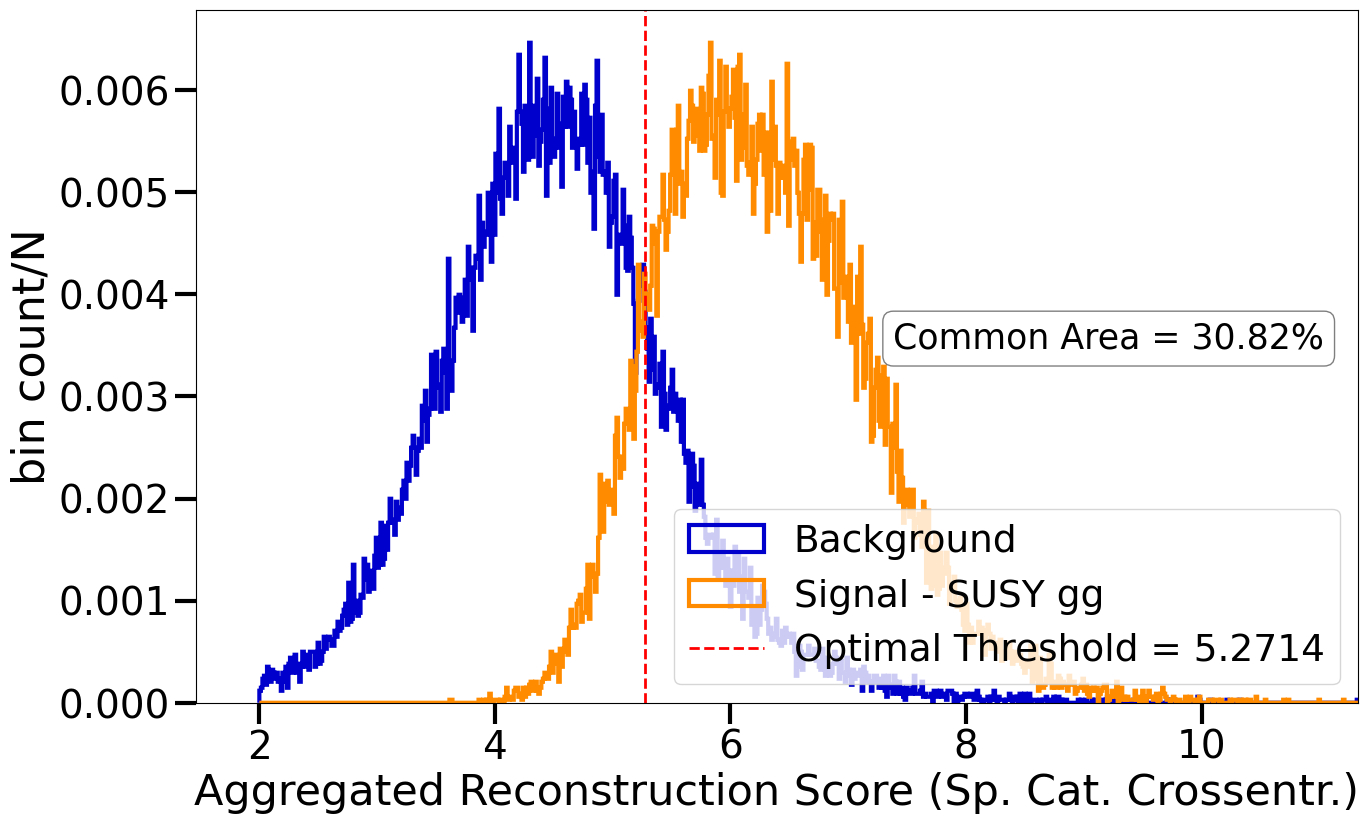} 
\caption{VQ-VAE tokenization, vocabulary size = 850.}
\label{fig:Histogram850-BSM}
\end{subfigure}

\vspace{0.5em}

\begin{subfigure}{0.49\textwidth}
\includegraphics[width=\linewidth]{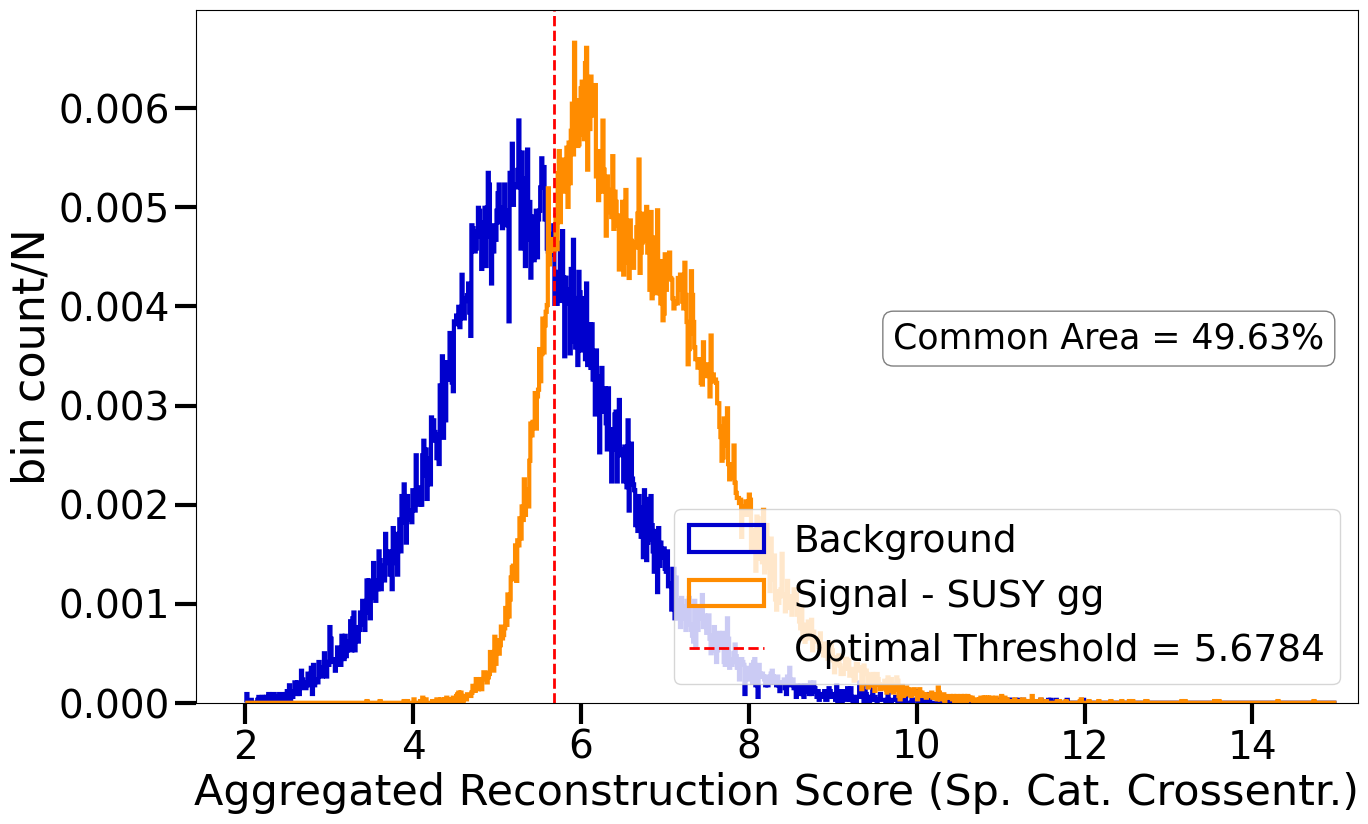} 
\caption{LUT tokenization, vocabulary size = 1524.}
\label{fig:Histogram1524-BSM}
\end{subfigure}
\hfill
\begin{subfigure}{0.49\textwidth}
\includegraphics[width=\linewidth]{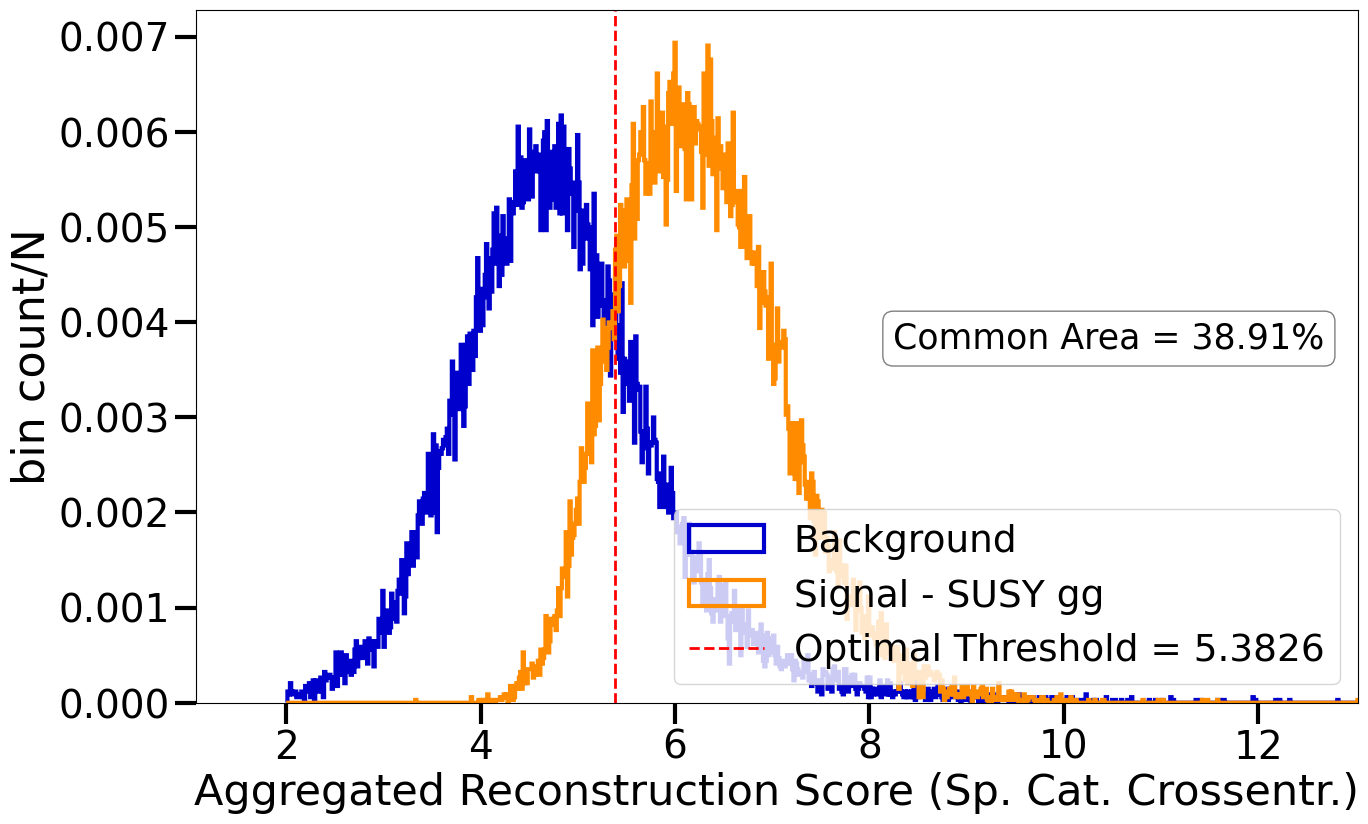} 
\caption{VQ-VAE tokenization, vocabulary size = 1700.}
\label{fig:Histogram1700-BSM}
\end{subfigure}
\caption{Anomaly score distributions for background (blue) and signal (orange) events in the SUSY $\tilde{g}\tilde{g}$ benchmark, as obtained from the downstream model. The optimal discrimination threshold is shown in red.}
\label{fig:histoBSM}
\end{figure}

\newpage

\clearpage
\bibliographystyle{JHEP}
\bibliography{refs.bib}

\end{document}